\documentclass[final,5p,twocolumn]{elsarticle}

\usepackage[english]{babel}
\usepackage{graphics}                 % Packages to allow inclusion of graphics
\usepackage{graphicx}
\usepackage[cmex10]{amsmath}
\usepackage{amssymb, amsfonts}
\usepackage{algorithmic}
\usepackage{algorithm}
\usepackage{verbatim}
\usepackage{url}
\usepackage{subfigure}
\usepackage{slashbox}
\usepackage{setspace}
\usepackage[numbers]{natbib}
\usepackage[hang,small]{caption}

\graphicspath{{figures/}}
\begin{document}

\begin{frontmatter}
\title{Tensor-Based Link Prediction in Intermittently Connected Wireless Networks}

\author{Mohamed-Haykel ZAYANI}
\ead{mohamed-haykel.zayani@telecom-sudparis.eu}

\author{Vincent GAUTHIER}
\ead{vincent.gauthier@telecom-sudparis.eu}

\author{Ines SLAMA}
\ead{ines.slama@telecom-sudparis.eu}

\author{Djamal ZEGHLACHE}
\ead{djamal.zeghlache@telecom-sudparis.eu}

\address{Lab. CNRS SAMOVAR UMR 5157,
Telecom SudParis, Evry, France}

\begin{abstract}
Through several studies, it has been highlighted that mobility
patterns in mobile networks are driven by human behaviors. This
effect has been particularly observed in intermittently connected
networks like DTN (Delay Tolerant Networks). Given that common
social intentions generate similar human behavior, it is relevant to
exploit this knowledge in the network protocols design, e.g. to
identify the closeness degree between two nodes. In this paper, we
propose a temporal link prediction technique for DTN which
quantifies the behavior similarity between each pair of nodes and
makes use of it to predict future links. Our prediction method keeps
track of the spatio-temporal aspects of nodes behaviors organized as
a third-order tensor that aims to records the evolution of the
network topology. After collapsing the tensor information, we
compute the degree of similarity for each pair of nodes using the
Katz measure. This metric gives us an indication on the link
occurrence between two nodes relying on their closeness. We show the
efficiency of this method by applying it on three mobility traces:
two real traces and one synthetic trace. Through several
simulations, we demonstrate the effectiveness of the technique
regarding another approach based on a similarity metric used in DTN.
The validity of this method is proven when the computation of score
is made in a distributed way (i.e. with local information). We
attest that the tensor-based technique is effective for temporal
link prediction applied to the intermittently connected networks.
Furthermore, we think that this technique can go beyond the realm of
DTN and we believe this can be further applied on every case of
figure in which there is a need to derive the underlying social
structure of a network of mobile users.

% we adapt a link prediction technique inspired
%from data-mining. This technique, inspired from the data-mining context, is based on two major steps. Firstly, the network structure is
%recorded over $T$ periods through several adjacency matrices that form a tensor. Secondly, . Katz scores . This closeness strongly privilege the link occurrence for the period $T$+1.  during the
%$T$ periods.
 %it is interesting to measure

\end{abstract}

\end{frontmatter}

\section{Introduction}
In recent years, extensive research has addressed challenges and
problems raised in mobile, sparse and intermittently connected
networks (i.e. DTN). In this case, forwarding packets tightly
depends on contacts occurrence. Since the existence of links is
crucial to deliver data from a source to a destination, the contacts
and their properties emerge as a key issue in designing efficient
communication protocols \cite{Hossmann2010a}. Obviously, the
occurrence of links is led by the behavior of the nodes in the
network \cite{Chaintreau07}. It has been widely shown in
\cite{Hsu2009a, Thakur2010} that human mobility is directed by
social intentions and reflects spatio-temporal regularity. A node
can follow other nodes to a specific location (spatial level) and
may bring out a behavior which may be regulated by a schedule
(temporal level). The social intentions that govern the behavior of
mobile users have also been observed through statistical analyses in
\cite{Chaintreau07,Karagiannis2007} by showing that the distribution
of inter-contact times follow truncated power law.

With the intention of improving the performance of intermittently
connected wireless network protocols, it is paramount to track and
understand the behaviors of the nodes. We aim at proposing an
approach that analyzes the network statistics, quantifies the social
relationship between each pair of nodes and exploits this measure as
a score which indicates if a link would occur in the immediate
future. We strongly believe that the social ties between nodes
highly govern the status of a link and establishes an indication for
the link prediction: it would never occur if two nodes have no
common social interactions and would rather be effective and lasting
with more correlated moving patterns.

In this paper, we adapt a tensor-based link prediction algorithm
successfully designed for the data-mining context
\cite{Acar2009,Dunlavy2011}. Our proposal records the network
structure for $T$ time periods and predicts links occurrences for
the $($T$+1)^{th}$ period. This link prediction technique is
designed through two steps. First, tracking time-dependent network
snapshots in adjacency matrices which form a tensor. Second,
applying of the Katz measure \cite{Katz1953} inspired from
sociometry. The link prediction technique computes the degree of
behavior similarity of each pair of nodes relying on the tensor
obtained in the first step. A high degree of behavior similarity
means that the two nodes have the same ``social" intentions. These
common intentions are expressed by the willingness to meet each
other and/or by similar moving patterns to visit a same location.
They also promote the link occurrence between two socially close
nodes in the immediate future (prediction of the period $T$+1 after
tracking the behaviors of nodes during $T$ time periods).

We further discuss how we design the tensor-based prediction method
and detail the two main steps in order to achieve link prediction.
On the one hand, we describe how to track the network topology over
time with a tensor. On the other hand, we explain how to compute and
interpret the Katz measure. We then evaluate the effectiveness of
predictability through several simulation scenarios depending on the
nature of the trace (real or synthetic), the number of recording
periods and the similarity metric computation which can be used in a
centralized or distributed way. Besides, to the best of our
knowledge, this work is the first to perform the prediction
technique in a distributed way. The assessment of its efficiency can
be beneficial for the improvement or the design of communication
protocols in mobile, sparse and intermittently connected networks.

The paper is organized as follows: Section 2 presents the related
work that highlights the growing interest to the social analysis and
justifies the recourse to the tensors and to the Katz measure to
perform predictions. In Section 3, we emphasize the two main steps
that characterize our proposal. Section 4 details simulation
scenarios used to evaluate the tensor-based prediction approach,
analyzes the obtained results, assesses its efficiency and proposes
a discussion about the described link prediction technique. Finally,
we conclude the paper in Section 5.

%%%%%%%%%%%%%%%%%%%%%%%%%%%%%%%%%%%%%%%%%%%%%%%%%%%%%%%%%%%%%%%%%%%%%%%%%%%%%%%%%%%%%%
%
% Section 2
%
%%%%%%%%%%%%%%%%%%%%%%%%%%%%%%%%%%%%%%%%%%%%%%%%%%%%%%%%%%%%%%%%%%%%%%%%%%%%%%%%%%%%%%
\section{Related Work}
The Social Network Analysis (SNA) \cite{Wasserman1994,
Katsaros2010a} and ad-hoc networking have provided new perspectives
for the design of network protocols \cite{Hui2008, Daly2007,
Hossmann2010}. These protocols aim to exploit the social aspects and
relationship features between the nodes. Studies conducted in the
field of SNA have mainly focused on two kinds of concepts: the most
well-known centrality metrics suggested in
\cite{Wasserman1994,Page1999,Hwang2008,Chung1997} and the community
detection mechanisms proposed in
\cite{Bollobas1998,Newman2006,Palla2005,Wasserman1994}. From this
perspective, several works have tried to develop synthetic models
that aim to reproduce realistic moving patterns
\cite{Hsu2009a,Lee2009}. Nonetheless, the study done in
\cite{Hossmann2010a} has outlined that synthetic models cannot
faithfully reproduce the human behavior because these synthetic
models are only location-driven and they do not track social
intentions explicitly. We consider in this work the Time-Variant
Community mobility model (TVC model) \cite{Hsu2009a}. The TVC model
depends on two main characteristics that influence the behavior of
nodes: geographical location preferences and time-dependent
behavior. This design tries to be closer to human-based behavior and
implicitly reproduces the social aspects that characterizes ad-hoc
networks.

Nevertheless, \cite{Katsaros2010a} has underlined the limits of
these protocols when the network topology is time-varying. The main
drawback comes down to their inability to model topology changes as
they are based on graph theory tools. Nevertheless, the tensor-based
approaches have been used in some works to build statistics on the
behaviors of nodes in wireless networks over time as in
\cite{Acer2010}. Thakur et al. \cite{Thakur2010} have also developed
a model using a collapsed tensor that tracks user's location
preferences (characterized by probabilities) with a considered time
granularity (week days for example) in order to considered the
emergence of ``behavior-aware" delay tolerant networks.

In this paper, we propose a link prediction technique that tracks
the temporal network topology evolution in a tensor and computes a
metric in order to characterize the social-based behavior similarity
of each pair of nodes. Some approaches have addressed the same
problem in data-mining in order to perform link prediction. Acar et
al. \cite{Acar2009} and Dunlavy et al. \cite{Dunlavy2011} have
provided detailed methods based on matrix and tensor factorizations
for link prediction in social networks such as the DBLP data set
\cite{DBLP}. These methods have been successfully applied to predict
a collaboration between two authors relying on the data set of the
structures of relationships over time. Moreover, they have
highlighted the use of the Katz measure \cite{Katz1953}, which can
be seen as a similarity metric, by assigning a link prediction score
for each pair of nodes. The efficiency of the Katz measure in link
prediction has been also demonstrated in
\cite{Wang2007,Liben-Nowell2007}.

\section{Description of the Tensor Based Prediction Method}
It has been highlighted that a human mobility pattern shows a high
degree of temporal and spatial regularity, and each individual is
characterized by a time-dependent mobility pattern and a trend to
return to preferred locations \cite{Chaintreau07, Hsu2009a,
Thakur2010}. In order to improve the design of wireless network
protocols, and especially the intermittently connected networks, it
is important to exploit this knowledge since these interactions
usually have an impact on the network structure and consequently on
the network performance. Thus, in this paper, we propose an approach
that aims to exploit similar behavior of nodes in order to predict
link occurrence referring to the social closeness.

Predicting future links based on their social closeness is a
challenge that is worth an investigation. Indeed, a good link
prediction technique contributes to improve the opportunistic
forwarding of packets and also enhances the delivery rate and/or
decreases latency. Moreover, it helps to avoid situations where
packets encumber the queue of the nodes that are not able to forward
them towards their final destinations.

To quantify the social closeness between each pair of nodes in the
network, we use the Katz measure \cite{Katz1953} inspired from
sociometry. This measure aims at measuring the social distance
between persons inside a social network. We also need to use a
structure that records link occurrence between each pair of nodes
over a certain period of time in order to perform the similarity
measure computation. The records represent the network behavior
statistics in time and space. To this end, tensors are used. A
tensor $\boldsymbol{\mathcal{Z}}$ consists in a set of slices and
each slice corresponds to an adjacency matrix of the network tracked
over a given period of time $D$. After the tracking phase, we reduce
the tensor into a matrix (or collapsed tensor) which expresses the
weight of each link according to its lifetime and its recentness. A
high weight value in this matrix denotes a link whose corresponding
nodes share an important degree of closeness. We apply the Katz
measure on the collapsed tensor to compute a matrix of scores
$\mathbf{S}$ that not only considers direct links but also indirect
links (multi-hop connections). The matrix of scores expresses the
degree of similarity of each pair of nodes respecting to the spatial
and the temporal levels. The higher the score is, the better the
similarity pattern gets. Therefore, two nodes that have a high
similarity score are most likely expected to have a common link in
the future.

\subsection{Notation}
Scalars are denoted by lowercase letters, e.g., $a$. Vectors are
denoted by boldface lowercase letters, e.g., $\bf{a}$. Matrices are
denoted by boldface capital letters, e.g., $\mathbf{A}$. The
$r^{th}$ column of a matrix $\mathbf{A}$ is denoted by $\bf{a_r}$.
Higher-order tensors are denoted by bold Euler script letters, e.g.,
$\boldsymbol{\mathcal{T}}$. The $n^{th}$ frontal slice of a tensor
$\boldsymbol{\mathcal{T}}$ is denoted $\mathbf{T_n}$. The $i^{th}$
entry of a vector $\bf{a}$ is denoted by $\bf{a}(i)$, element
$(i,j)$ of a matrix $\mathbf{A}$ is denoted by $\mathbf{A}(i,j)$,
and element $(i, j, k)$ of a third-order tensor
$\boldsymbol{\mathcal{T}}$ is denoted by $\mathbf{T_{i}}(j, k)$.

\subsection{Matrix of Scores Computation}
\begin{figure*}[!tb]
    \centering
    \includegraphics[width=0.9\textwidth]{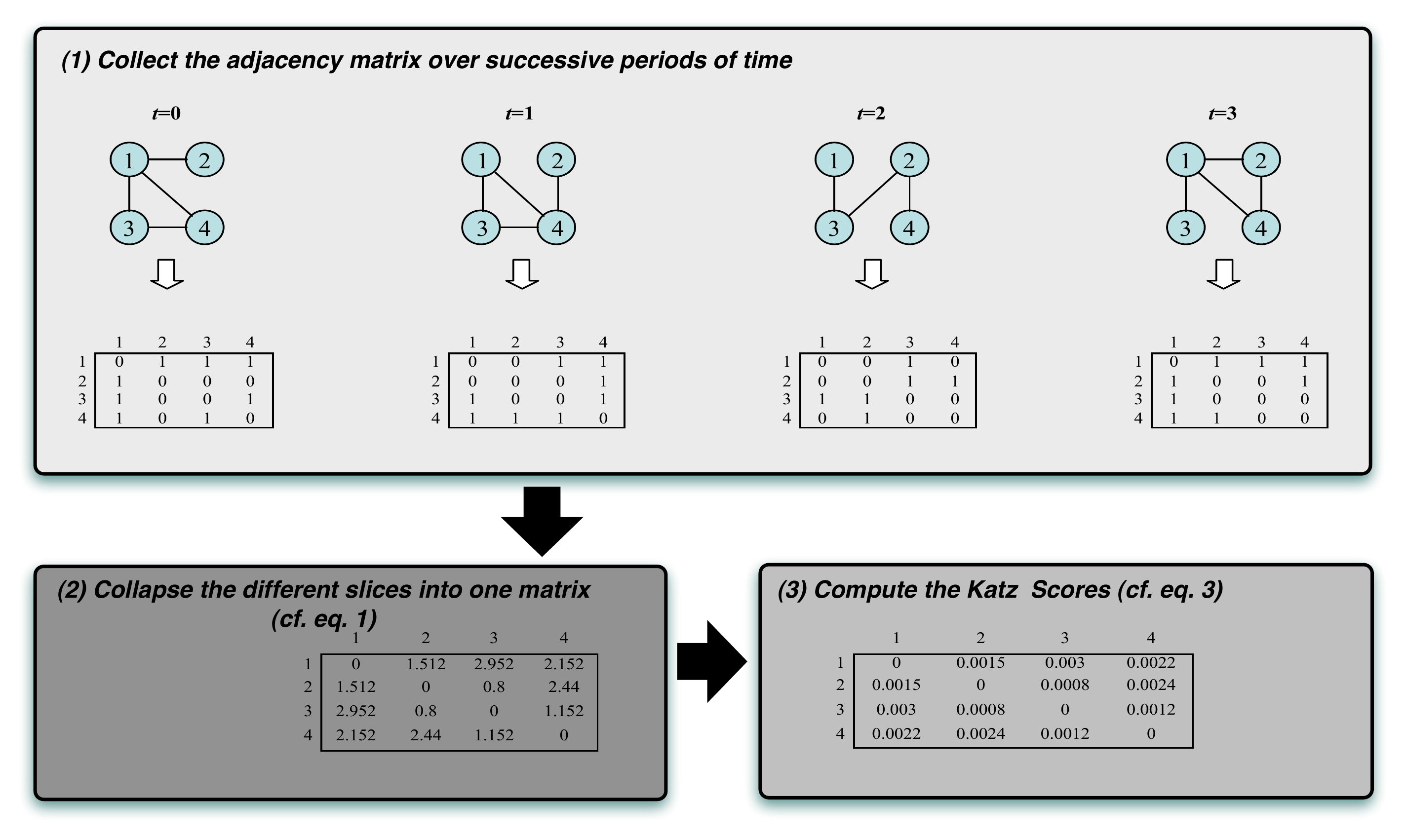}
    \caption{Example of the matrix $\mathcal{S}$ computation}
    \label{Zayani0}
\end{figure*}

The computation of the similarity scores is modeled through two
distinct steps. First, we store the inter-contact between nodes in a
tensor $\boldsymbol{\mathcal{Z}}$ and reduce it to a matrix
$\mathbf{X}$ called the collapsed tensor. In a second step, we
compute the matrix of similarity scores $\mathbf{S}$ relying on the
matrix $\mathbf{X}$ (cf. Fig. \ref{Zayani0}).

\subsubsection{Collapsing the data from the tensor}
We consider that the data is collected into the tensor
$\boldsymbol{\mathcal{Z}}$. The slice $\mathbf{Z_{t}}(i, j)$
describes the status of a link between a node $i$ and a node $j$
during a time period between $[tD,(t+1)D[$ where $\mathbf{Z}_{t}(i,
j)$ is 1 if the link exists during the time period $D$ and 0
otherwise. The tensor is formed by a succession of adjacency
matrices $\mathbf{Z_{1}}$ to $\mathbf{Z_{T}}$ where the subscript
letters designs the observed period. To collapse the data into one
matrix as done in \cite{Acar2009,Dunlavy2011}, we choose to compute
the collapsed weighted tensor (which is more efficient than
collapsed tensor as shown in \cite{Acar2009} and
\cite{Dunlavy2011}). The links structure is considered over time and
the more recent the adjacency matrix is, the more weighted the
structure gets.

\begin{equation}
    \mathbf{X}(i,j)=\sum_{t=0}^{T} (1-\theta)^{T-t}\ \mathbf{Z_{t}}(i,j)
    \label{eq1}
\end{equation}

Where the matrix $\mathbf{X}$ is the collapsed weighted tensor of
$\boldsymbol{\mathcal{Z}}$, and $\theta$ is a parameter used to
adjust the weight of recentness and is between 0 and 1.

\subsubsection{Katz Measure}
The Katz measure, which is affiliated to sociometry, was first
proposed by Leo Katz in \cite{Katz1953}. He considers a social
network as a undirected graph $G = (V,E)$ where each vertices
$V=\{v_1, v_2,...,v_k\}$ is a finite set of node that represent a
persons and each edge $E=\{e_1, e_2,...,e_k\}$ is a finite set of
connection (or an endorsement) between two persons. We denote a
subset $P_{\ell}(v_i,v_j) = \{e_1, e_2, ..., e_{\ell}\}\subset E$ as
a path of length $\ell$ between node $v_i$ and $v_j$. The score that
characterizes the couple $(v_i,v_j)$ is defined by the weight of
paths $P_{\ell}(v_i,v_j)$ connecting person $v_i$ to person $v_j,
\forall v \in V$.

Katz defined his metric between two nodes as $\mathbf{S}(i,j)$ as
depicted in Eq.\eqref{eq2}. It is a function that decreases
proportionally to the path length $P_{\ell}(v_i,v_j)$. Katz did so
in order to emphasize the fact that endorsements strength fade over
a successive chains of recommendations. This measure can be seen as
a generalization of followers (as in Twitter) or an indegree
measure. It indicates the number of votes of a person as well as the
identity of the voter (his vote is valuable compared to the number
of votes he receives). This metric is widely used in studies which
aim to predict links occurrence \cite{Acar2009, Dunlavy2011,
Liben-Nowell2007}, especially in social networks as co-authorship
communities as the DBLP \cite{DBLP} and arXiv \cite{arXiv}
databases. Given that there are ``social relationships" between
nodes in networks with intermittent connections, it is challenging
to exploit this measure and to apply it on collected data.
Therefore, the Katz score of a link between a node $i$ and a node
$j$ as given by \cite{Katz1953}:

\begin{equation}
    \mathbf{S}(i,j)=\sum_{\ell=1}^{+\infty} \beta^{\ell} P_{\ell}(v_i,v_j)
    \label{eq2}
\end{equation}

Where $\beta$ is a user defined parameter strictly superior to zero
and $\beta^{\ell}$ is the weight of a $\ell$ hops path length. It is
clear that the longer the path is, the lower the weight gets. There
is also another formulation to compute Katz scores by means of
collapsed weighted tensor as detailed previously. Then, the score
matrix $\mathbf{S}$ can be rewritten as:

\begin{equation}
    \mathbf{S}=\sum_{\ell=1}^{+\infty} \beta^{\ell} \cdot \mathbf{X}^{\ell}=(\mathbf{I}-\beta \cdot \mathbf{X})^{-1}-\mathbf{I}
    \label{eq3}
\end{equation}

Where $\mathbf{I}$ is the identity matrix and $\mathbf{X}$ is the
obtained collapsed weighted tensor.

In Fig. \ref{Zayani0}, we provide an example that describes the two
main steps of the link prediction technique. We consider a network
of 4 nodes whose topology is dynamic over time. At each period $t$
(from 0 to 3), each occurred link is caught in the corresponding
adjacency matrix. All adjacency matrices form the tensor. The latter
structure is used to determine the collapsed weighted tensor by
computing Eq. (\ref{eq1}) (by setting $\theta$ to 0.2) for each pair
of nodes. Then, the matrix of scores is computed by applying Eq.
(\ref{eq3}) ($\beta$ is set to 0.001) on the collapsed weighted
tensor.

The measure goes beyond estimating a link weight between two nodes.
Indeed, it takes into consideration all possible paths between two
nodes and then quantifies the social relationship between them. As
described previously, when two nodes are connected through short
paths, the score characterizing this pair is high. Hence, the score
can be treated as the node's moving pattern similarity, in view of
the fact that the nodes conserve their vicinity (short paths). When
two nodes share high score, this means that their behaviors are
similar and that they are geographically quite close. Therefore, a
link occurrence between them is very likely.

\subsection{Matrix of Scores Interpretation}
The relationship between each pair of nodes is expressed by a score
$\mathbf{S}(i,j)$, this score reflects the degree of similarity
between node $i$ and node $j$. As mentioned in the Katz measure
analysis, shorter paths lead to higher scores. Thus, two nodes that
share a high score are nodes that are connected through short paths
during some period of time and therefore have similar behaviors
(similar social intentions). The similarity here is related to
common preferences in spatial and temporal space. Two nodes maintain
their connectivity when they move in the same direction and at the
same time. Therefore, these scores can be considered as indicators
to a possible link existence in the future. Thus, the link
prediction is done through measuring behavior similarity for each
nodes pairs in the matrix $\mathbf{S}$.

The computation of matrix $\mathbf{S}$, as described before, is done
in a \textbf{centralized way}. It means that the matrix $\mathbf{S}$
is computed based on a full knowledge of the network topology over
time. This may not be suitable with ad hoc wireless networks where
no central entity is considered and could in addition be very
costly. A \textbf{distributed mechanism} should then be examined. In
a distributed mechanism, each node would apply the prediction method
relying only on information related to its nearest neighbors. It is
paramount to remember that a Katz formulation gives more weight to
short paths and assigns low scores to long paths. Therefore, the
scores with neighbors located at few hops away should be sufficient
and strong enough compared to scores with further ones. This
hypothesis will be discussed in Section 3.

\section{Performance Evaluation and Simulation Results}
To evaluate how efficient is the tensor-based link prediction in
intermittently connected wireless networks, we consider three
different traces (two real traces and one synthetic trace). For each
scenario, we compute the corresponding scores matrix $\mathbf{S}$ as
described earlier and assess the efficiency of the link prediction
method through evaluation techniques. In the following, we firstly
present the traces used for the link prediction evaluation. Then, we
expose the corresponding results and analyze the effectiveness of
the prediction method.

\subsection{Simulation Traces}
We consider three traces to evaluate the link prediction approach.
Two of them are real traces and the third is synthetic. We exploit
them to construct the tensor by generating adjacency matrices (with
different time period $t$: 5, 10, 30 and 60 minutes). At each case,
we track the required statistics about nodes behavior within $T$
periods. We also consider the adjacency matrix corresponding to the
period $T$+1 as a benchmark to evaluate Katz scores matrix. We
detail, in the following, the used traces.
\begin{itemize}
\item \textbf{First Trace: Dartmouth Campus trace:}
We choose the trace of 01/05/06 \cite{Dartmouth} and construct the
tensor slices relying on SYSLOG traces between 8 a.m. and midday (4
hours). The number of nodes is 1018.
\item \textbf{Second Trace: MIT Campus trace:}
We focus on the trace of 07/23/02 \cite{Balazinska2003} and consider
also the events between 8 a.m. and midday to build up the tensor.
The number of nodes is 646.
\item \textbf{Third Trace: TVC Model trace:}
In this scenario, we use the trace generator proposed by Hsu et al.
\cite{TVCM} which reproduces the concept of the TVC model. We
consider a square simulation area with an edge length equal to 1000
meters and where 100 nodes are in motion. We randomly generate two
locations as the node's geographical preferences and keep community
switching and roaming probabilities as in the example provided in
the generator. As in the other scenarios, we track nodes behavior
during 4 hours. Table \ref{table1} summarizes the main parameters
considered in generating TVC Model traces.
\end{itemize}

\begin{table}[!t]
\renewcommand{\arraystretch}{1.3}
\caption{Major Parameters Used in TVC Model} \label{table1}
\centering \scalebox{0.7}{
\begin{tabular}{|c|c|}
\hline
\bfseries Parameter & \bfseries Value\\
\hline\hline
Simulation Area Edge Length & 1000 meters \\
Network Nodes Number & 100 nodes \\
Network Nodes Range & 75 meters \\
Network Geographical Communities Number & 2 \\
Maximum Nodes Speed & 15 m/s \\
Minimum Nodes Speed & 5 m/s \\
Average Nodes Speed & 10 m/s \\
 \hline
\end{tabular}}

\end{table}

For each scenario, we generate adjacency matrices corresponding to a
different period $t$: 5, 10, 30 and 60 minutes. Then, to record the
network statistics over 4 hours, the tensor has respectively a
number of slices $T$ equal to 48, 24, 8 and 4 slices (for the case
where $t$=5 minutes, it is necessary to have 48 periods to cover 4
hours). As mentioned earlier, we take into account both centralized
and distributed cases for the computation of scores.
\begin{itemize}
\item \textbf{The Centralized Computation:}
The centralized way assumes that there is a central entity which has
full knowledge of the network structure at each period and applies
Katz measure to the global adjacency matrices.
\item \textbf{The Distributed Computation:}
Each node has a limited knowledge of the network structure. We
assume that a node is aware of its two-hop neighborhood. Hence,
computation of Katz measures is performed on a
local-information-basis.
\end{itemize}
In both cases, we fix $\theta$ and $\beta$ to 0.2 et 0.001
respectively. Later, we explain why we choose these values.

\subsection{Performance Analysis}
As described in the previous section, we apply the link prediction
method to the three types of traces while considering the different
tensor slice periods in both centralized and distributed cases. In
order to assess the efficiency of this method, we consider several
link prediction scenarios (according to the trace, the tensor slice
period and the scores computation way) and we use different
evaluation techniques (ROC curves, CDF curves, AUC metric and top
scores ratio at $T$+1). We detail in the following the results
obtained with each evaluation technique and analyze the link
prediction efficiency.

\subsubsection{Analysis of the ROC Curves}
\begin{figure*}[!tb]
  \centering
  \subfigure[5 minutes tensor slice period]{
    \label{fig1}
    \includegraphics[width=0.25\textwidth,angle=270]{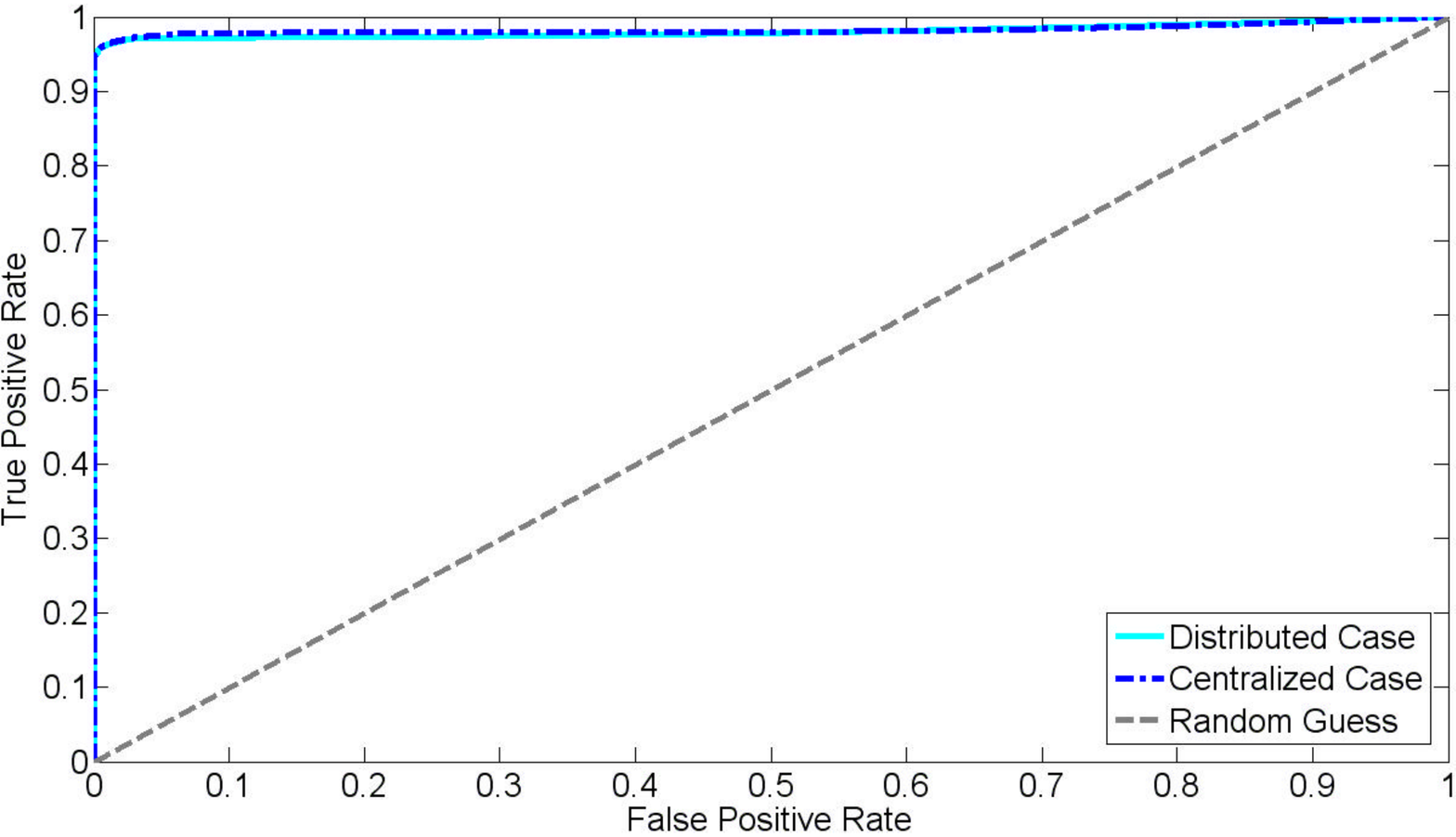}
  }\hspace{1cm}%\hfill
  \subfigure[10 minutes tensor slice period]{
    \label{fig2}
    \includegraphics[width=0.25\textwidth,angle=270]{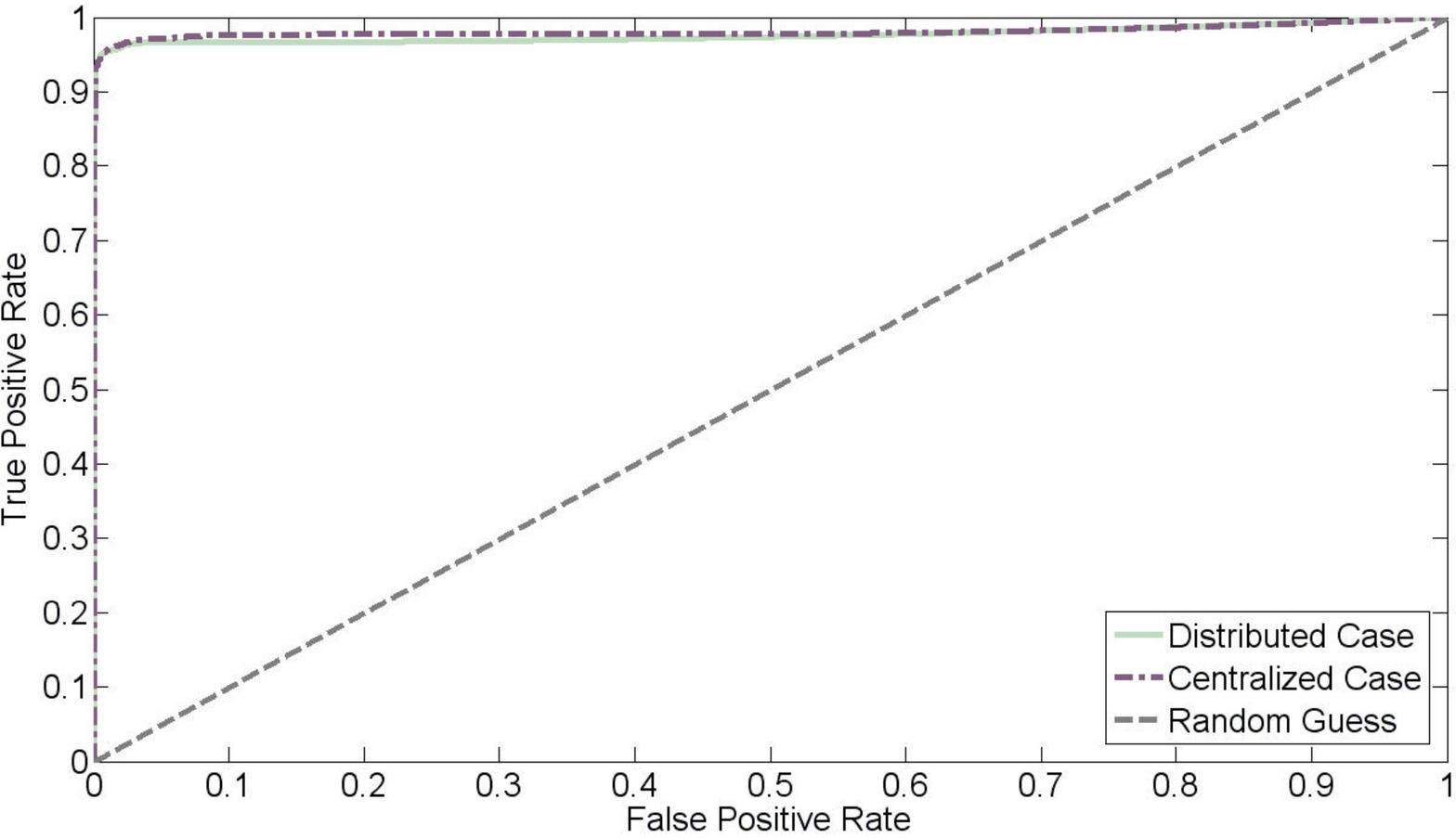}
  }\\
 \subfigure[30 minutes tensor slice period]{
    \label{fig3}
    \includegraphics[width=0.25\textwidth,angle=270]{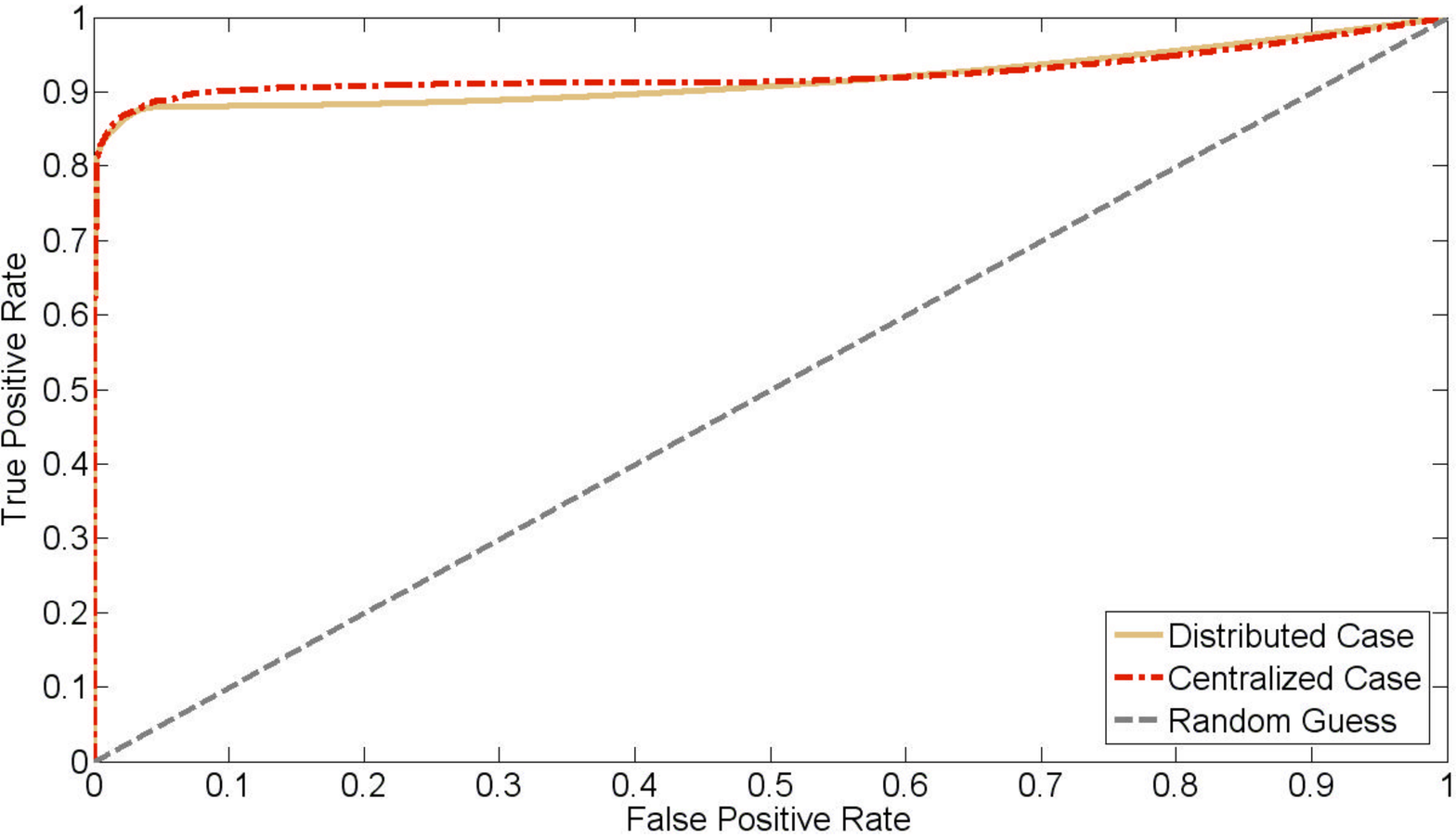}
  }\hspace{1cm}%\hfill
  \subfigure[60 minutes tensor slice period]{
     \label{fig4}
     \includegraphics[width=0.25\textwidth,angle=270]{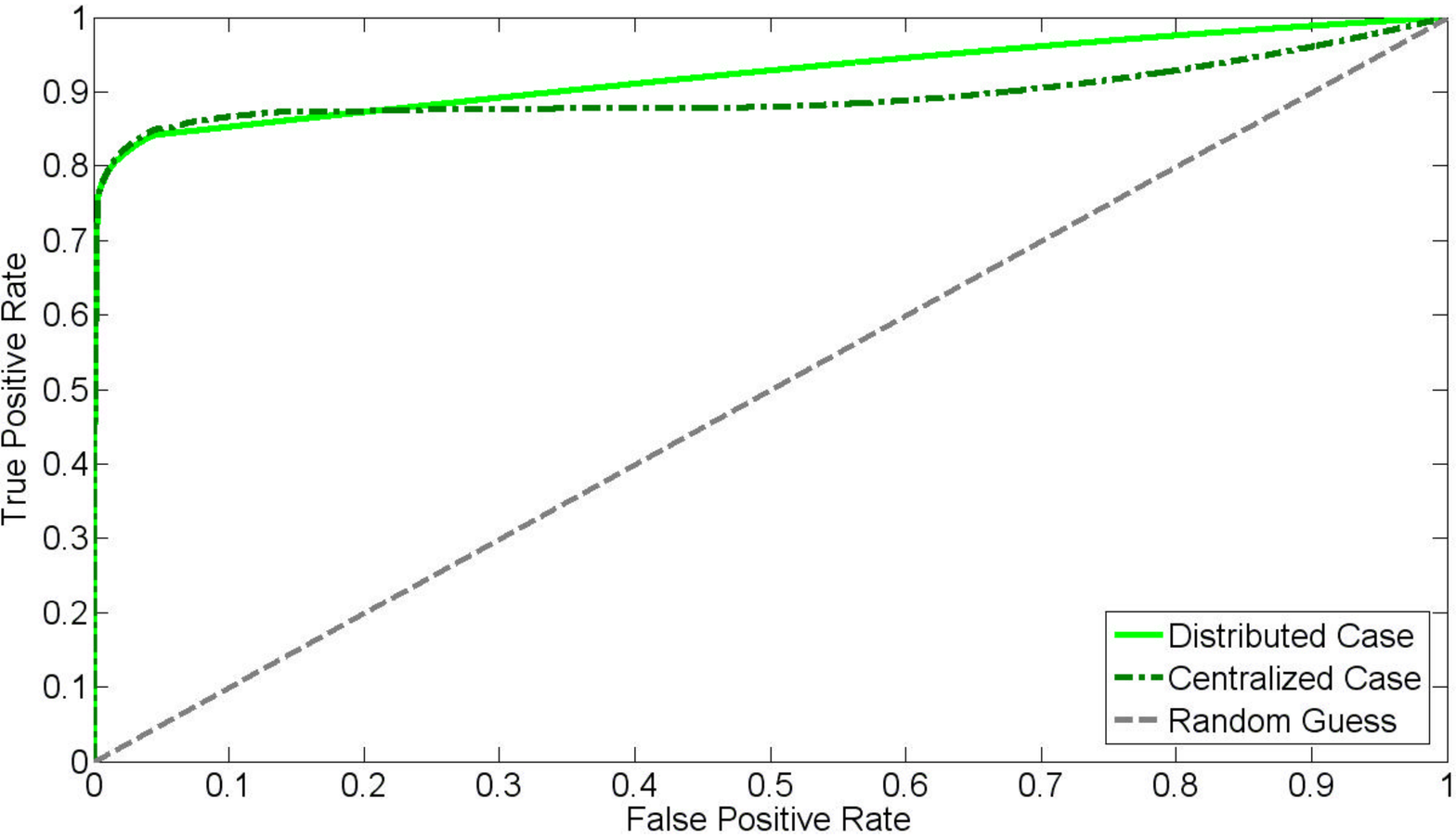}
  }
  \caption{ROC Curves for different prediction cases applied on Dartmouth Campus trace}
  \label{ROC_Dartmouth}
\end{figure*}

Fig. \ref{ROC_Dartmouth}, \ref{ROC_MIT} and \ref{ROC_TVC} depict the
ROC curves (Receiver Operating Characteristic) \cite{FAWCETT2006}
for both distributed and centralized computing approaches
respectively obtained from Dartmouth Campus trace, MIT Campus trace
and TVC model trace. For each trace figure, (a), (b), (c) and (d)
curves correspond to a tensor slice time of 5, 10, 30 and 60 minutes
respectively.

We first notice that, for all scenarios, the prediction of all links
is quite efficient, compared to the random guess (the curve's bends
are at the upper left corner). Moreover, two other observations have
to be mentioned in the case of real traces (Dartmouth Campus and MIT
Campus traces). First, it is highlighted that the smaller the tensor
slice (adjacency matrix) period is, the more reliable the prediction
gets. This observation is obvious for two reasons. On the one hand,
with a low tensor slice time, the probability of tracking a short
and occasional contact between two nodes is not likely. On the other
hand, recording four hours of statistics requires 48 adjacency
matrices of 5-minutes periods instead of 4 matrices for 60-minutes
periods case. Thus, tracking a short contact between two nodes has
less influence when the tensor slices are more numerous. As an
example, in the case where the tensor slice time is 5 minutes, a
fleeting contact can be caught by one adjacency matrix among 48.
However, for the case where the slice time is 60 minutes, the
fleeting contact is tracked by one tensor slice among 4, which
significantly gives it more weight compared to the former case.
Hence, short tensor slice periods enable us to minimize the
probability of tracking a short contact existence and to restrict
its impact.

Short tensor slice periods also allow us to better track the social
interactions (meetings in a cafeteria, courses in an amphitheater,
etc) between nodes which determine the occurrence of links.
Successive adjacency matrices of 5 minutes give more accurate
description of network structure over time as both analyzing and
identifying these social events are easier through smaller periods.

The second observation concerns the similar results obtained at the
centralized and distributed matrix of scores computation. In fact,
the similarity is higher when the paths considered between a pair of
nodes are short. Thereby, paths that have more than two hops have
weaker scores and so are less weighted compared to shorter ones. The
distributed case assumes that each node knows its neighbors at most
at two hops. That is why distributed scores computation presents
performances which are so similar to the centralized ones.
\begin{figure*}[!tb]
  \centering
  \subfigure[Distributed computation of scores]{
    \label{fig13}
    \includegraphics[width=0.25\textwidth,angle=270]{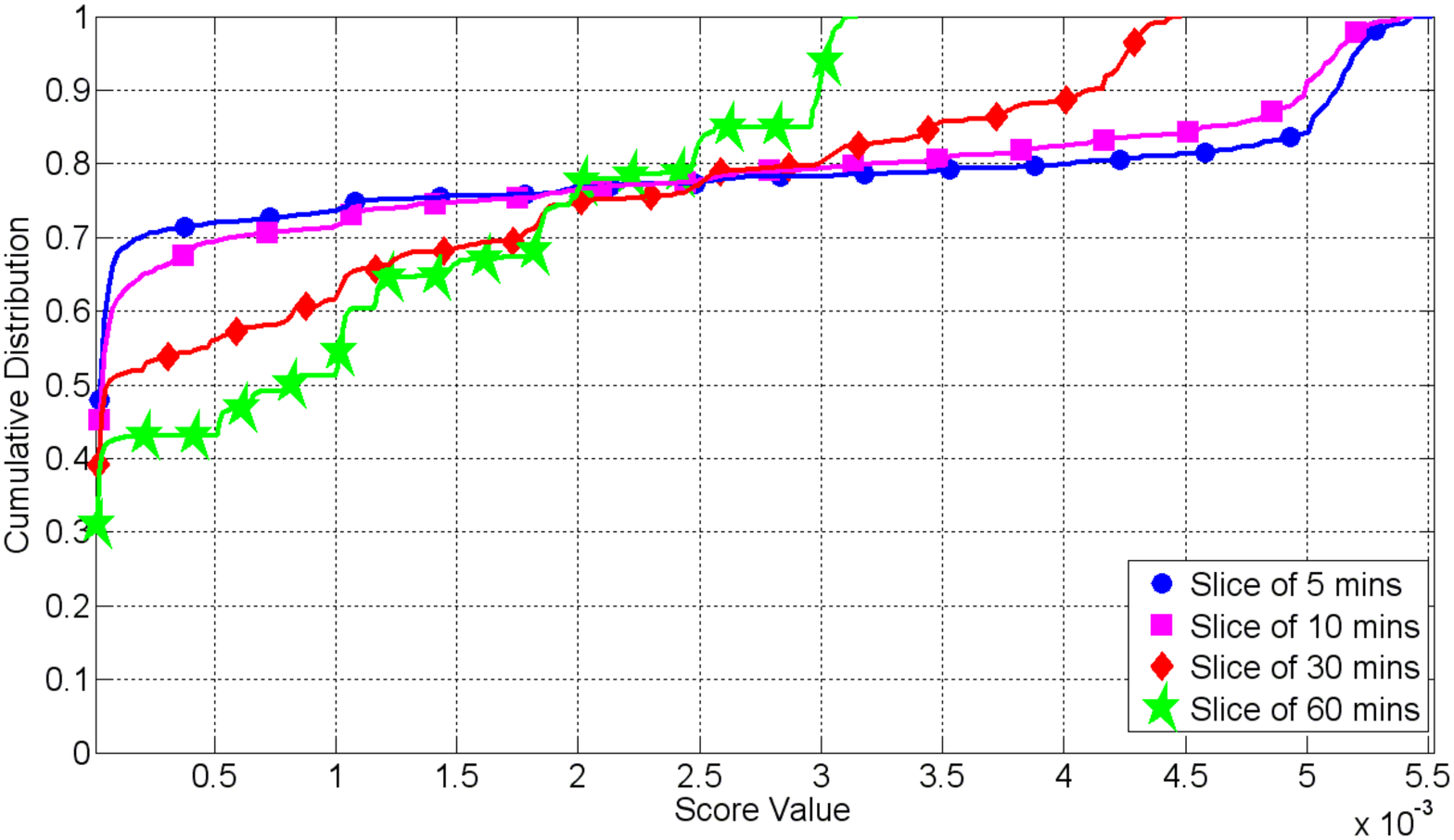}
  }\hspace{1cm}%\hfill
  \subfigure[Centralized computation of scores]{
    \label{fig14}
    \includegraphics[width=0.25\textwidth,angle=270]{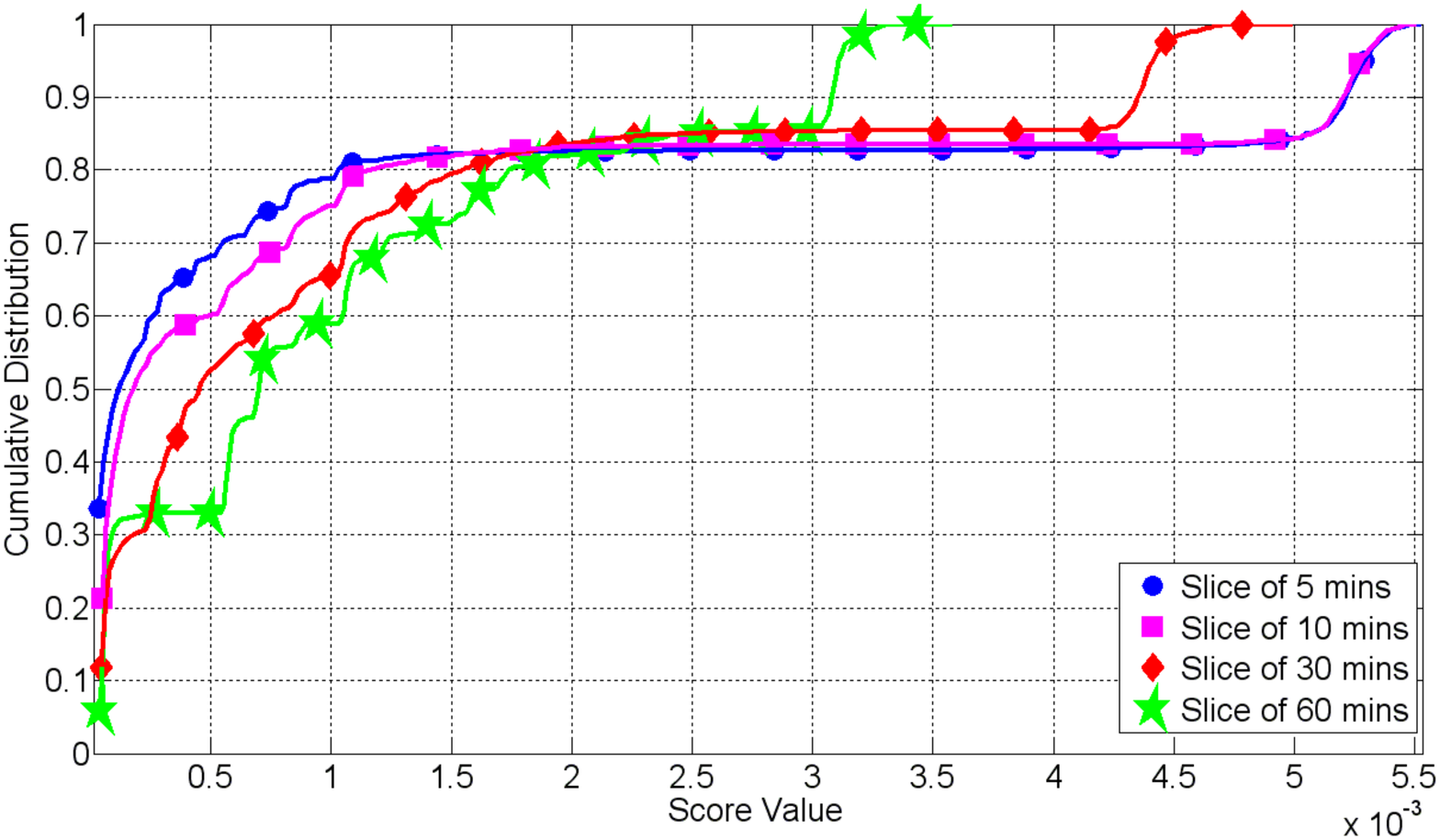}
  }
  \caption{Cumulative distribution function of Katz scores obtained from Dartmouth Campus trace}
  \label{CDF_Dartmouth}
\end{figure*}

Regarding the results obtained from the synthetic trace (TVC model
trace), it is obvious that there are no significant differences
between the ROC curves as the tensor slice periods varies
(especially for the scenarios where the period $t$ is higher than 5
minutes). On top of having the same performances with the two scores
matrix computation ways, changing the adjacency matrix time period
does not impact the link prediction efficiency. This observation
could be explained by the conclusion drawn by Hossmann et al. in
\cite{Hossmann2010a} which outlines that location-driven mobility
models do not care about social intentions. In addition, through the
proposed behavior similarity metric between a pair of nodes, Thakur
et al. \cite{Thakur2010} prove that TVC model limits moving patterns
to visiting preferred locations and do not take care of any social
coordination. With TVC model, movement patterns are the same for all
nodes (moving into two geographical communities) and repetitive (the
chosen moving speed is between 5 and 15 m/s with an average speed of
10 m/s). They are only regulated by geographical preferences (each
node visits the preferred community with an ``individual
willingness" and there is no correlation between its moving pattern
and those of other nodes). Therefore, having several tensor slices
is not different from considering fewer ones. Moreover, the
adjacency matrix at $T$+1 in each scenario is quite the same.

\subsubsection{Analysis of the CDF Curves}
In order to highlight the impact of the choice of the period $t$ on
the link prediction, we represent in Fig. \ref{CDF_Dartmouth},
\ref{CDF_MIT} and \ref{CDF_TVC} the skewed Cumulative Distribution
Function (CDF) of the scores obtained respectively for Dartmouth
Campus trace, MIT Campus trace and TVC model trace (only strictly
positive scores are considered). At each trace's CDF figure, (a) and
(b) correspond to the distributed and centralized scores matrix
computation respectively. The obtained results for real traces
(Dartmouth Campus and MIT Campus) show that the spreading of
distribution is narrower when the period $t$ is larger. In fact, at
a CDF with wider spreading (especially at the case of $t$=5 min),
high scores that express link occurrence prediction are easier to
figure out. On the contrary, the interval of scores is narrow and so
the score's analysis is more imprecise. These results confirm the
ones obtained through ROC curves. While the CDF results of real
traces look similar, the ones of the synthetic trace show that the
tensor slice period has a less significative impact. Indeed, the
cumulative distribution functions are redundant at over 80\% of
obtained scores (when the scores are situated between 0 and
$3.10^{-3}$). This observation also applies to ROC curves results.

As a final note, we underline that the synthetic trace CDF shows a
higher percentage of weak scores than those of real traces. This
observation explains the more limited prediction efficiency outlined
with the TVC model trace.

\begin{figure*}[!tb]
  \centering
  \subfigure[5 minutes tensor slice period]{
    \label{fig5}
    \includegraphics[width=0.25\textwidth,angle=270]{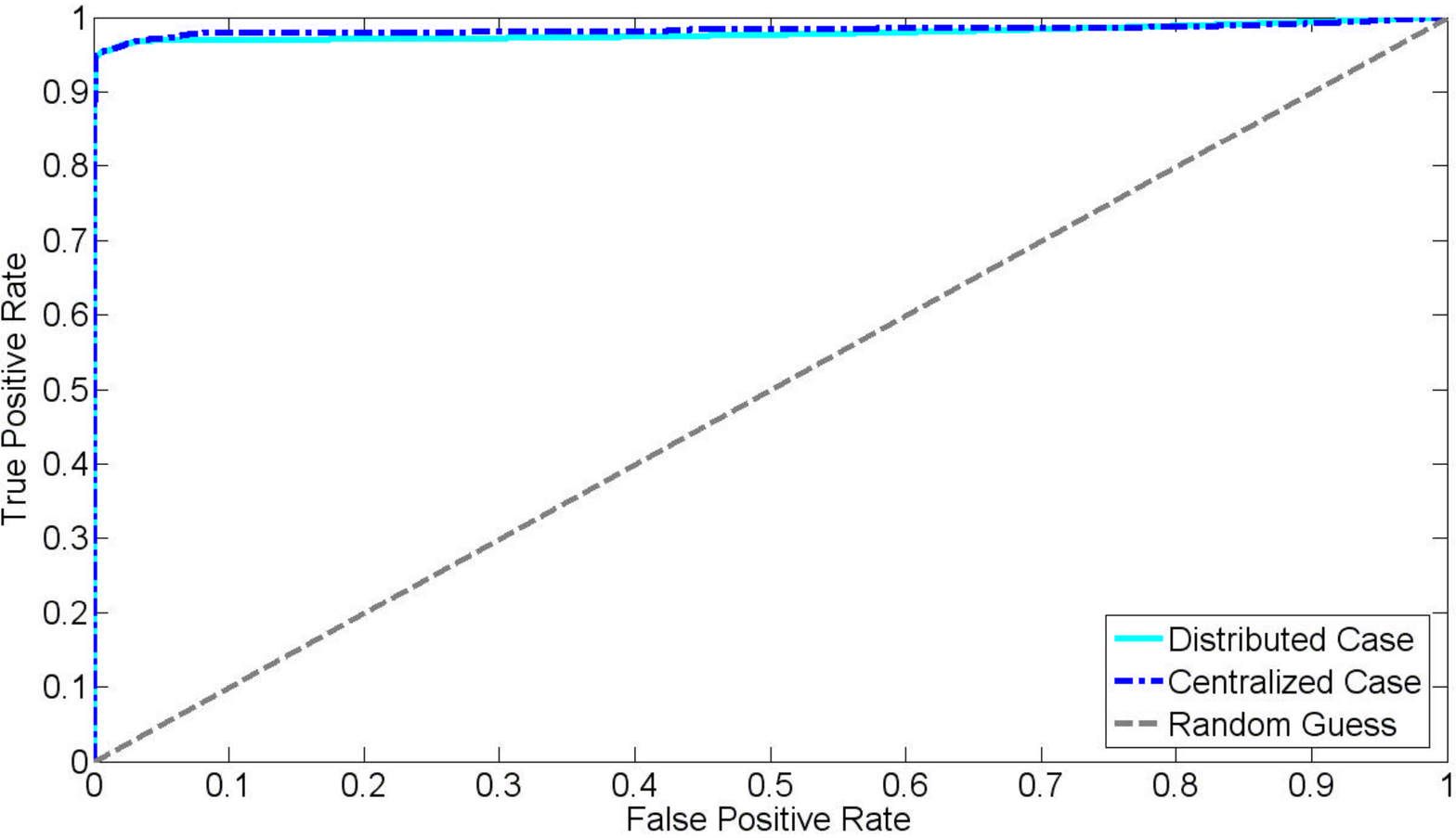}
  }\hspace{1cm}%\hfill
  \subfigure[10 minutes tensor slice period]{
    \label{fig6}
    \includegraphics[width=0.25\textwidth,angle=270]{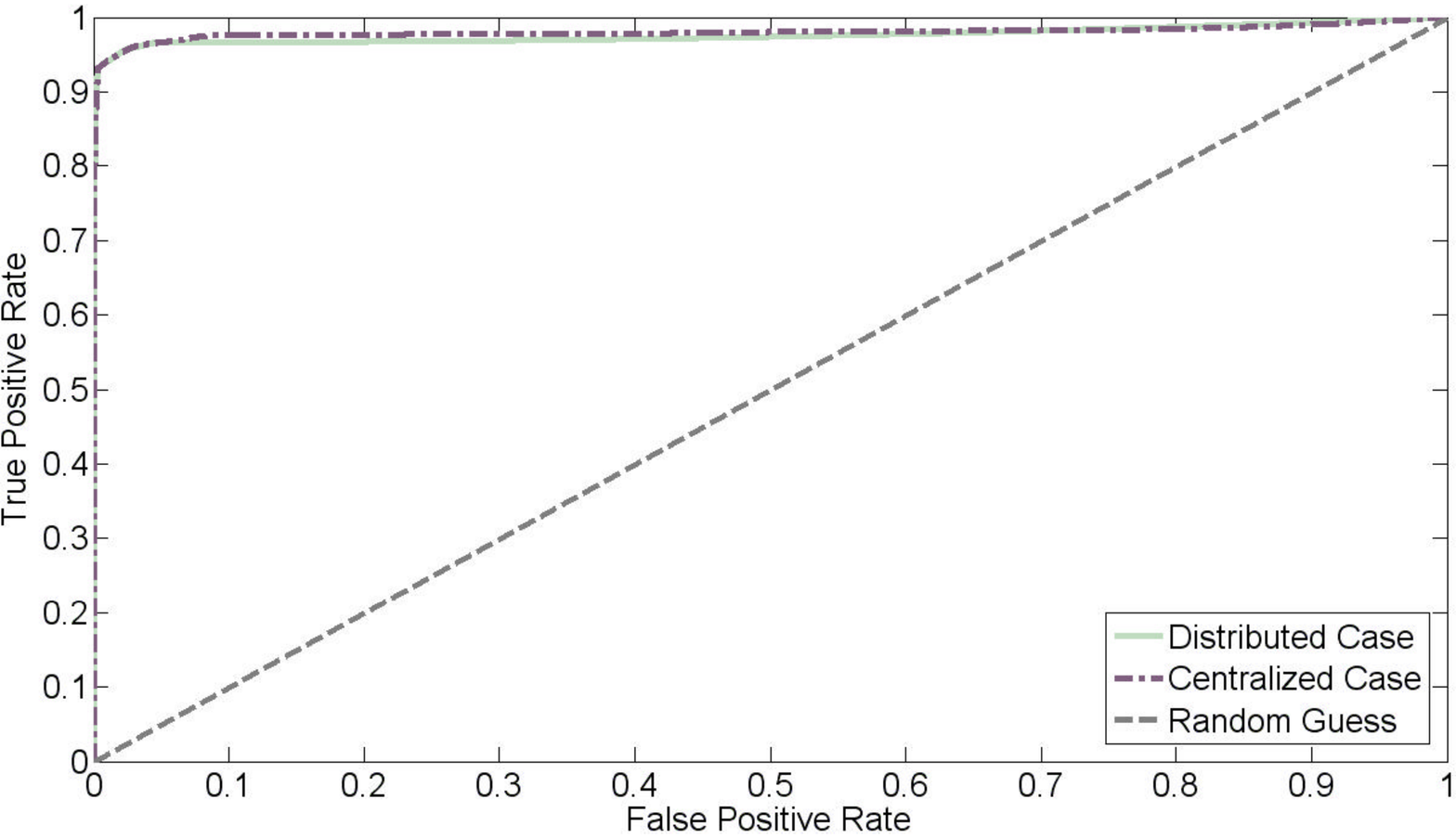}
  }\\
 \subfigure[30 minutes tensor slice period]{
    \label{fig7}
    \includegraphics[width=0.25\textwidth,angle=270]{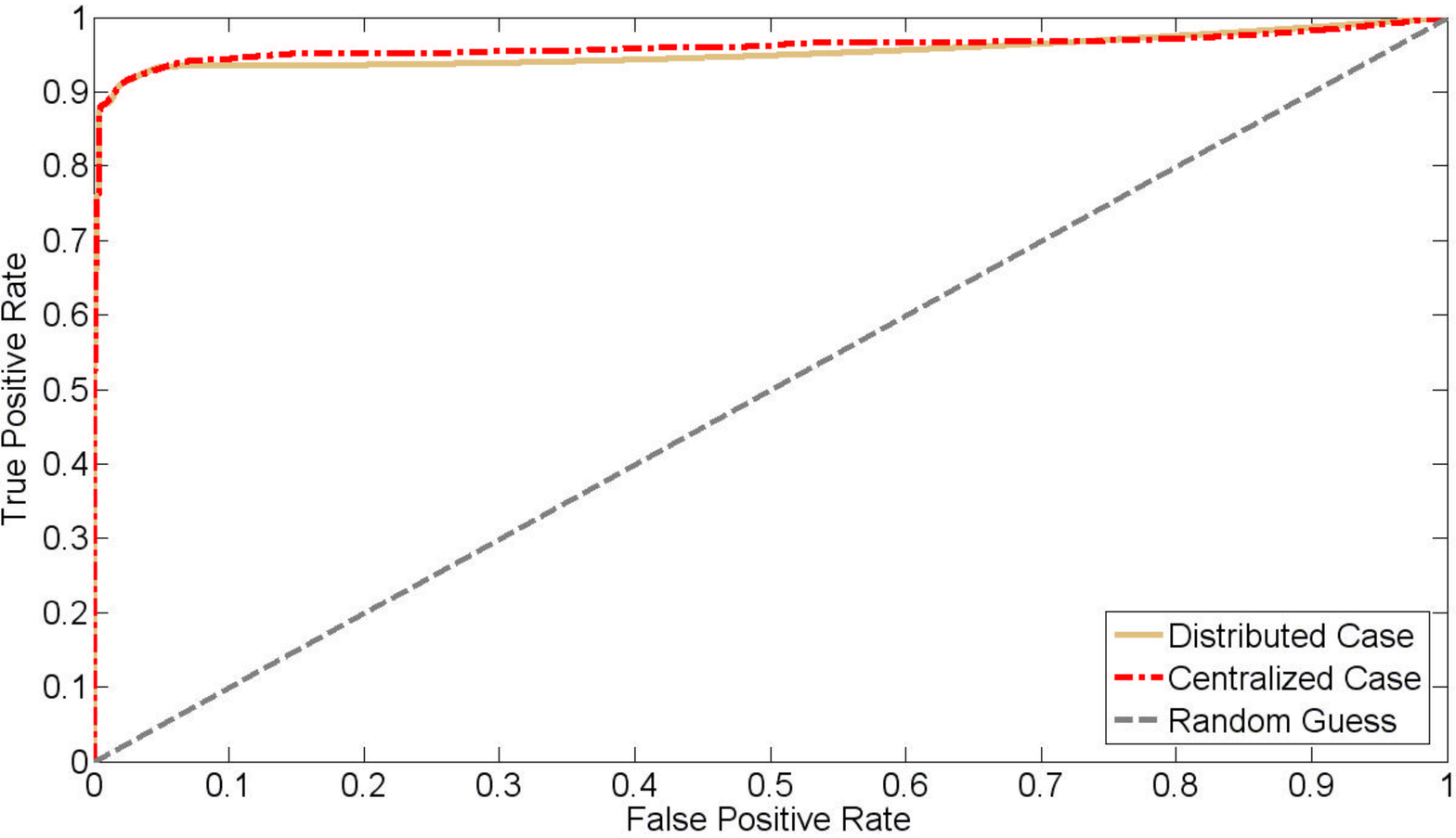}
  }\hspace{1cm}%\hfill
  \subfigure[60 minutes tensor slice period]{
     \label{fig8}
     \includegraphics[width=0.25\textwidth,angle=270]{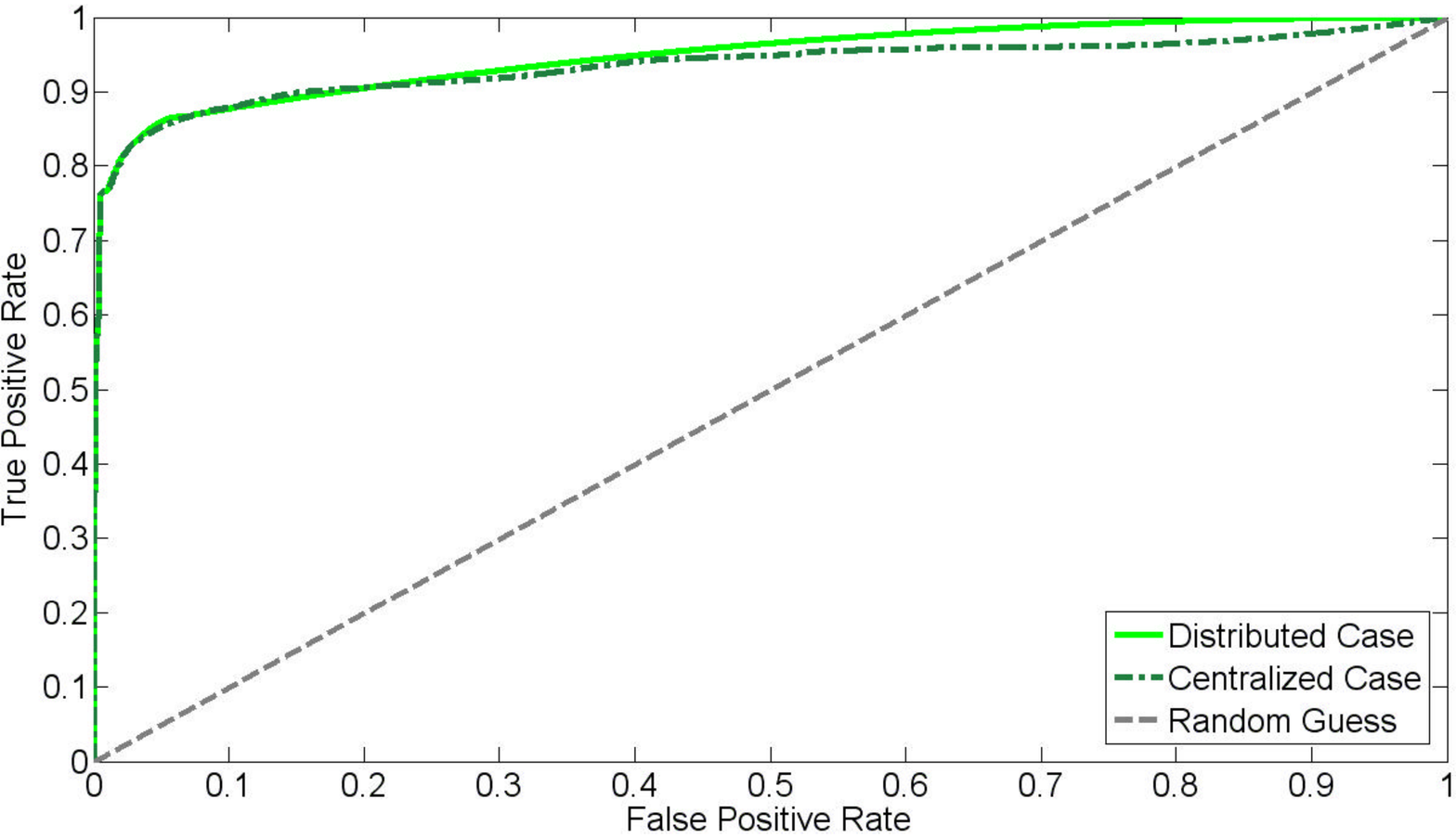}
  }
  \caption{ROC Curves for different prediction cases applied on MIT Campus trace}
  \label{ROC_MIT}
\end{figure*}
\begin{figure*}[!tb]
  \centering
  \subfigure[Distributed computation of scores]{
    \label{fig15}
    \includegraphics[width=0.25\textwidth,angle=270]{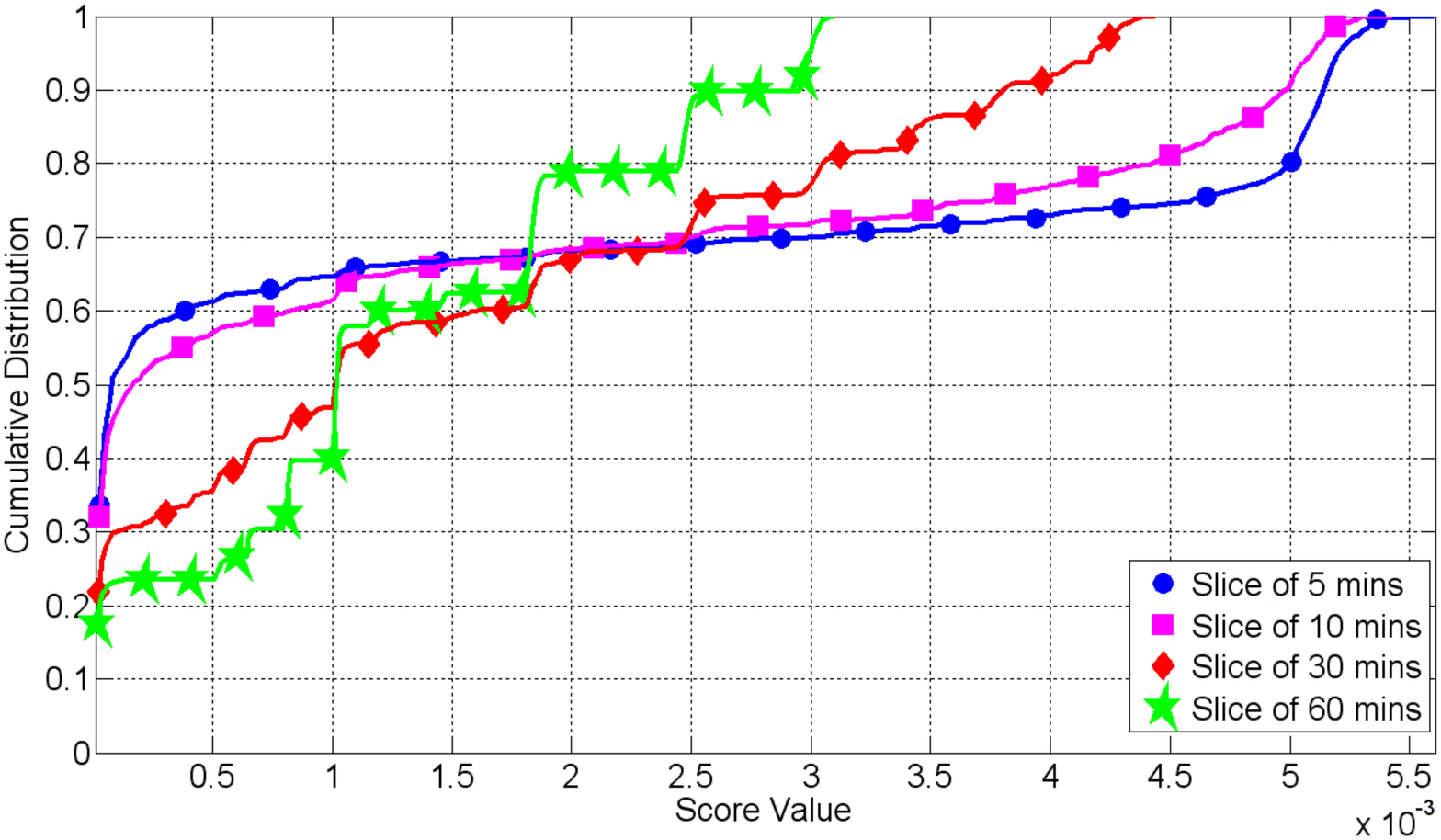}
  }\hspace{1cm}%\hfill
  \subfigure[Centralized computation of scores]{
    \label{fig16}
    \includegraphics[width=0.25\textwidth,angle=270]{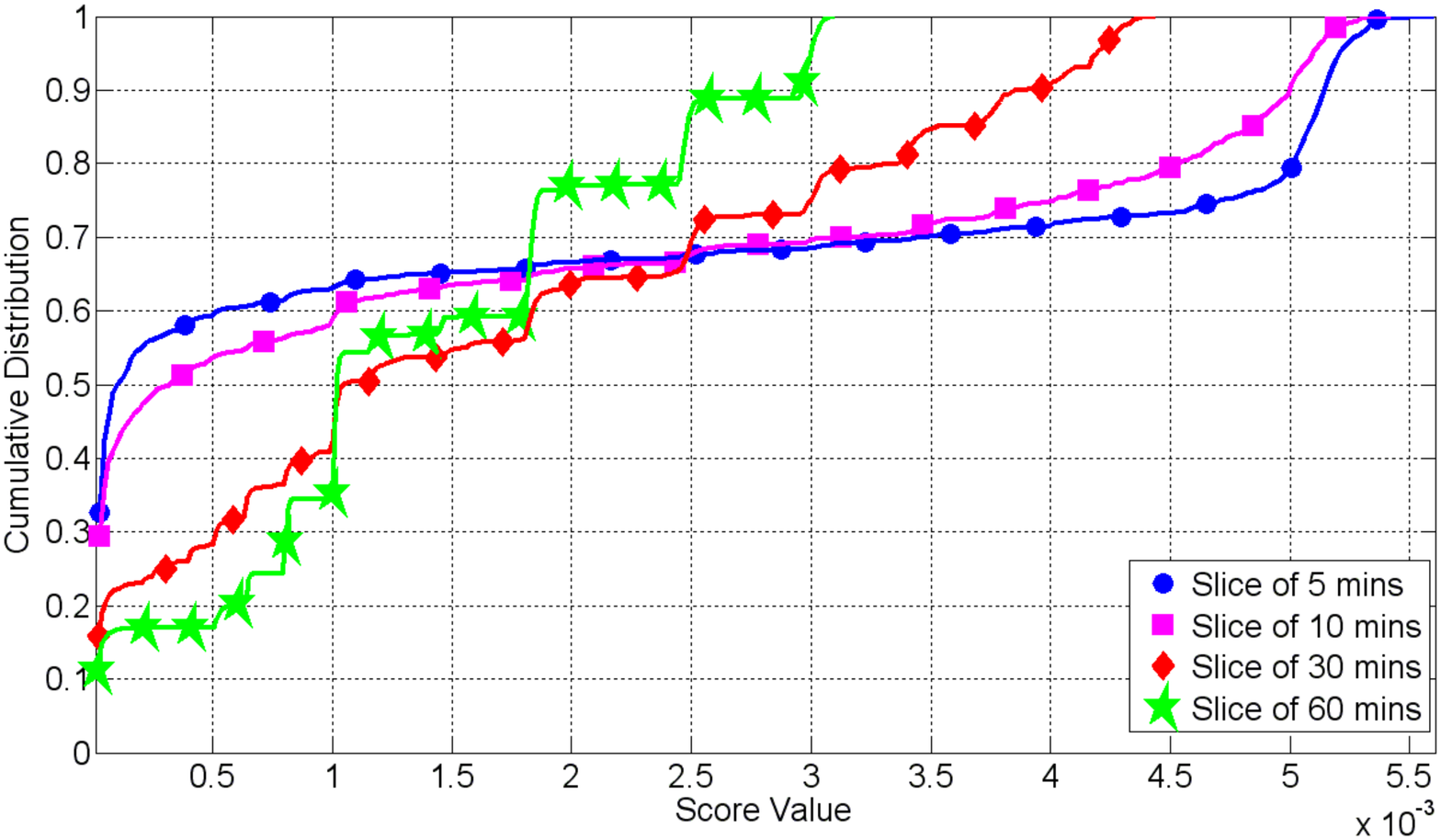}
  }
  \caption{Cumulative distribution function of Katz scores obtained from MIT Campus trace}
  \label{CDF_MIT}
\end{figure*}

\begin{figure*}[!tb]
  \centering
  \subfigure[5 minutes tensor slice period]{
    \label{fig9}
    \includegraphics[width=0.25\textwidth,angle=270]{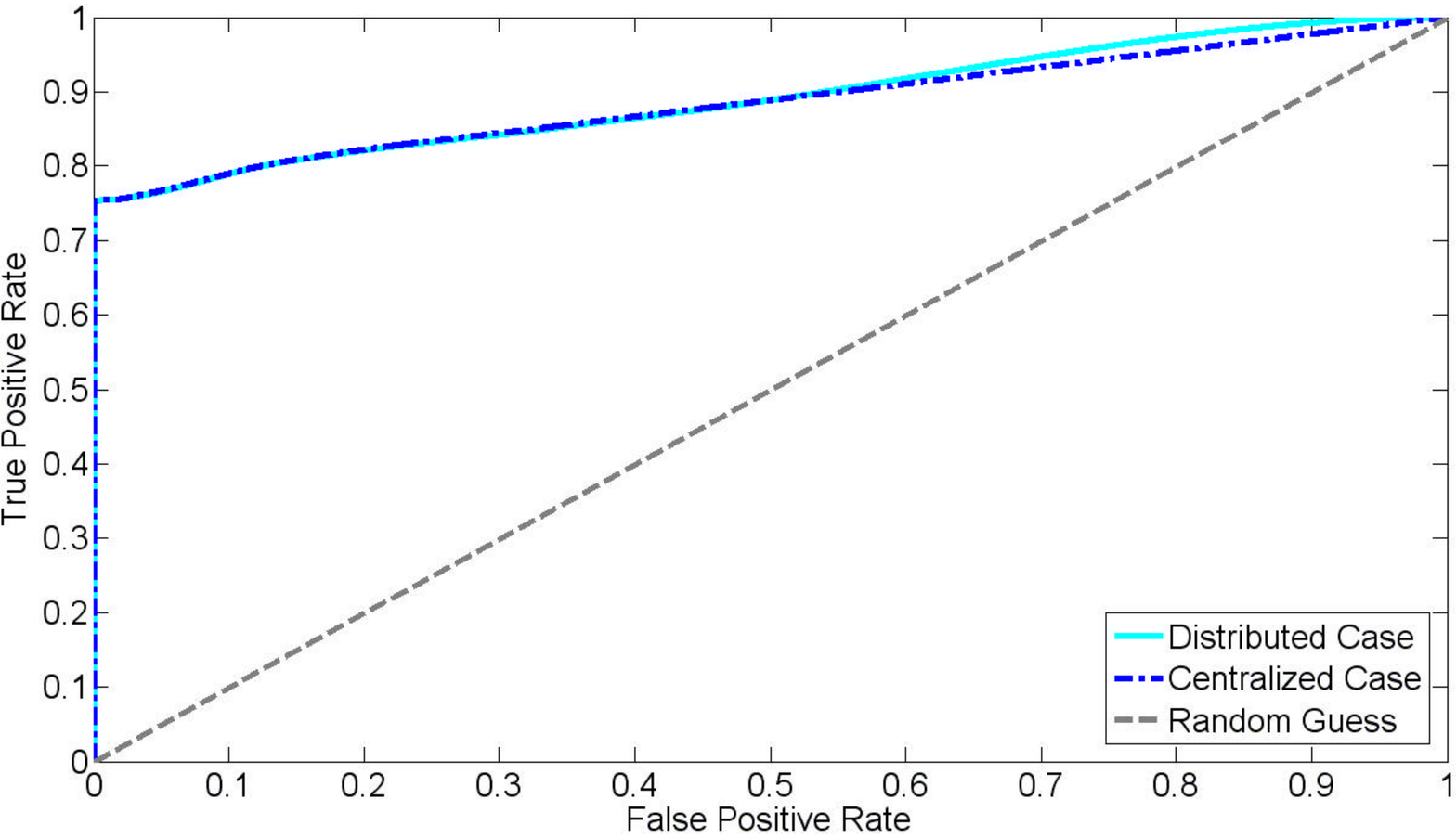}
  }\hspace{1cm}%\hfill
  \subfigure[10 minutes tensor slice period]{
    \label{fig10}
    \includegraphics[width=0.25\textwidth,angle=270]{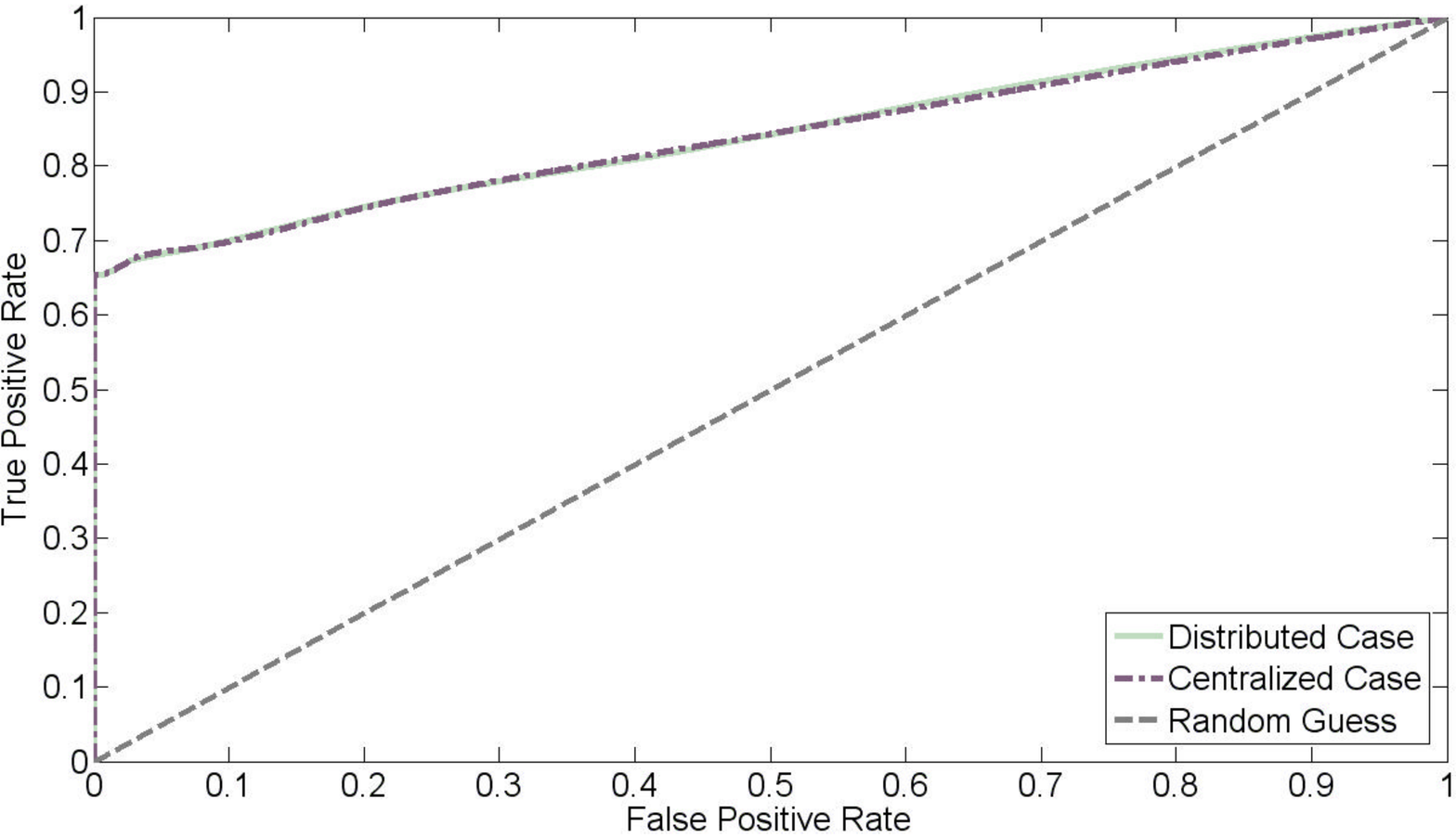}
  }\\
 \subfigure[30 minutes tensor slice period]{
    \label{fig11}
    \includegraphics[width=0.25\textwidth,angle=270]{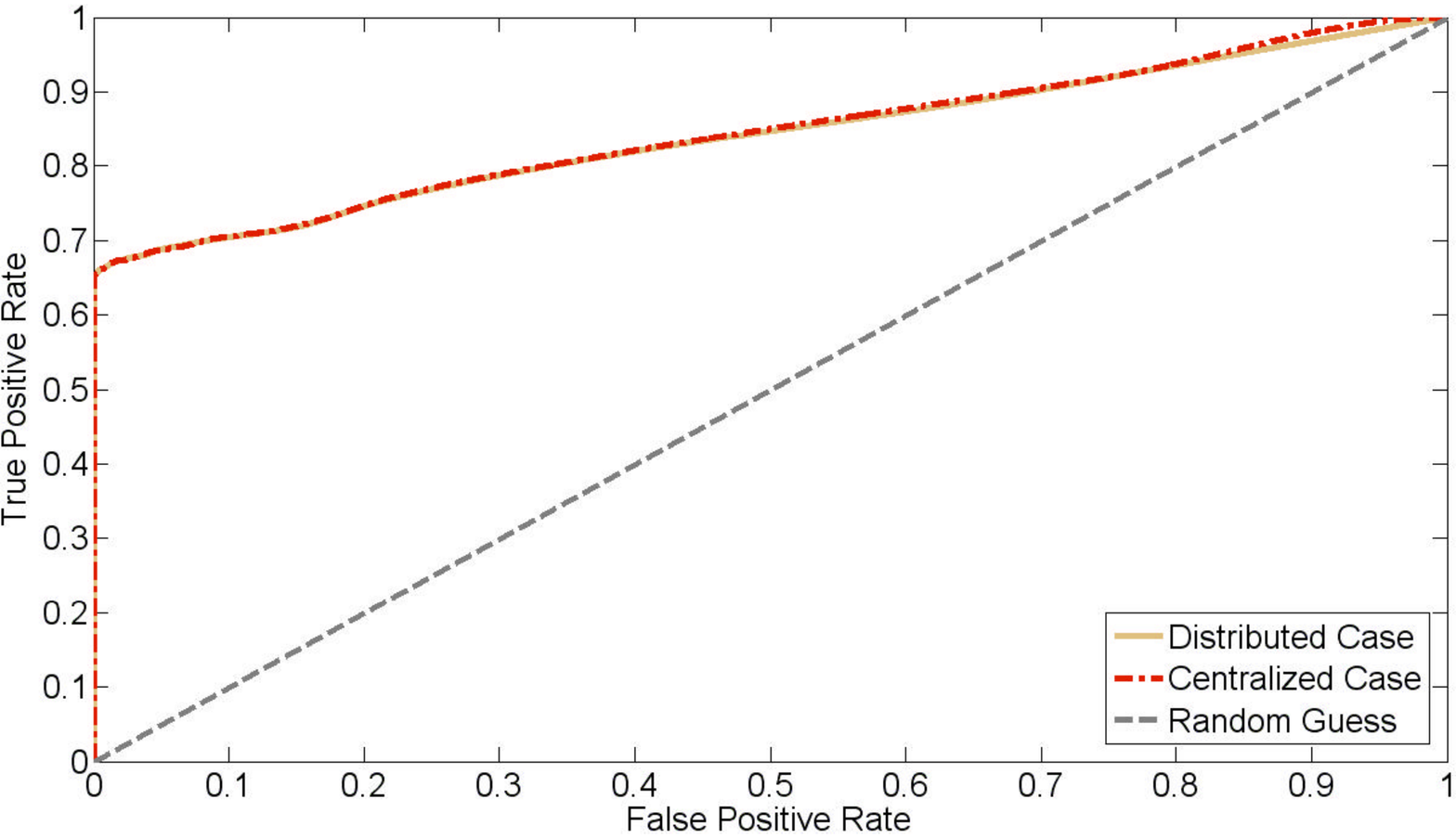}
  }\hspace{1cm}%\hfill
  \subfigure[60 minutes tensor slice period]{
     \label{fig12}
     \includegraphics[width=0.25\textwidth,angle=270]{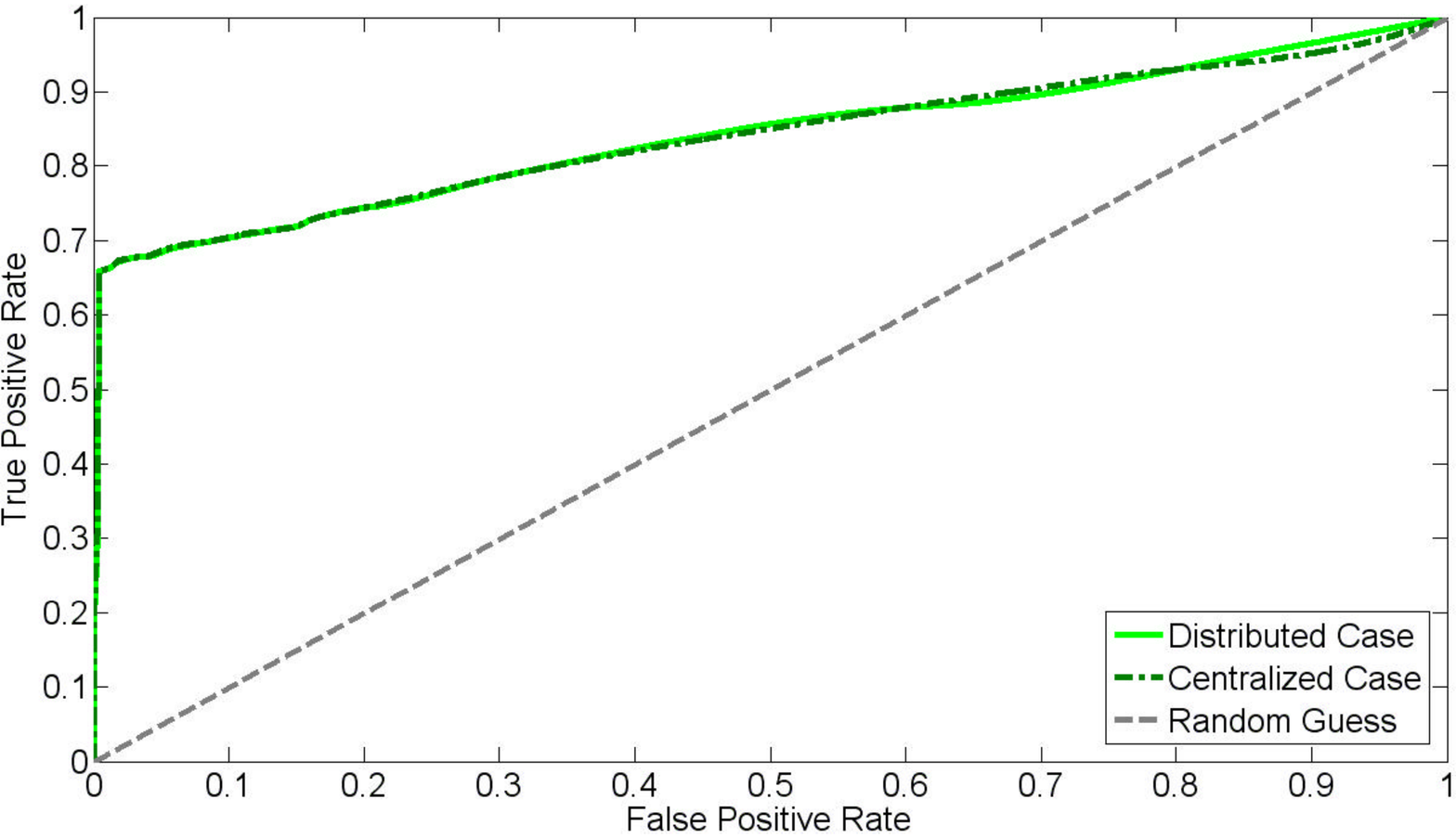}
  }
  \caption{ROC Curves for different prediction cases applied on TVC model trace}
  \label{ROC_TVC}
\end{figure*}
\begin{figure*}[!tb]
  \centering
  \subfigure[Distributed computation of scores]{
    \label{fig17}
    \includegraphics[width=0.25\textwidth,angle=270]{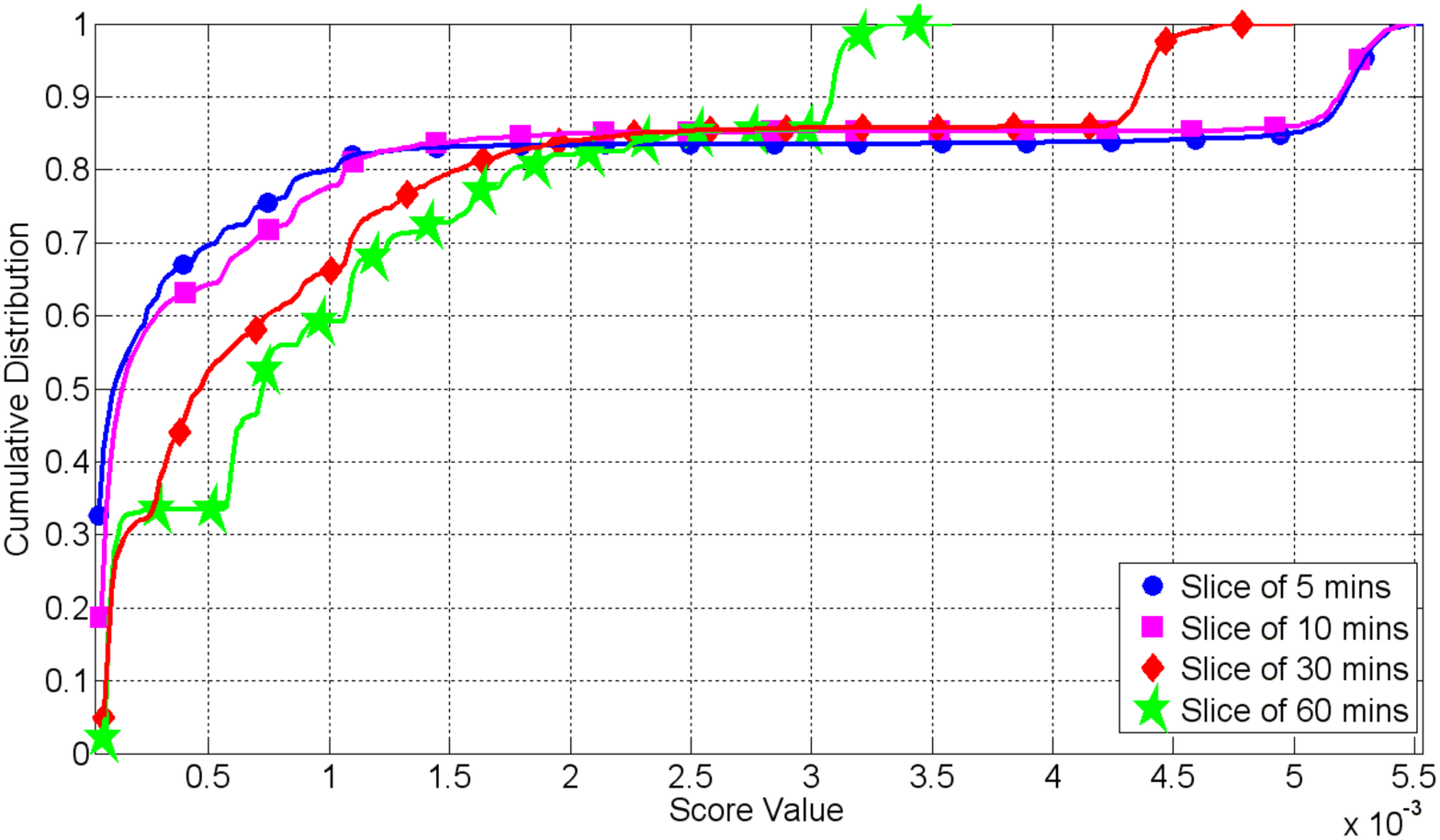}
  }\hspace{1cm}%\hfill
  \subfigure[Centralized computation of scores]{
    \label{fig18}
    \includegraphics[width=0.25\textwidth,angle=270]{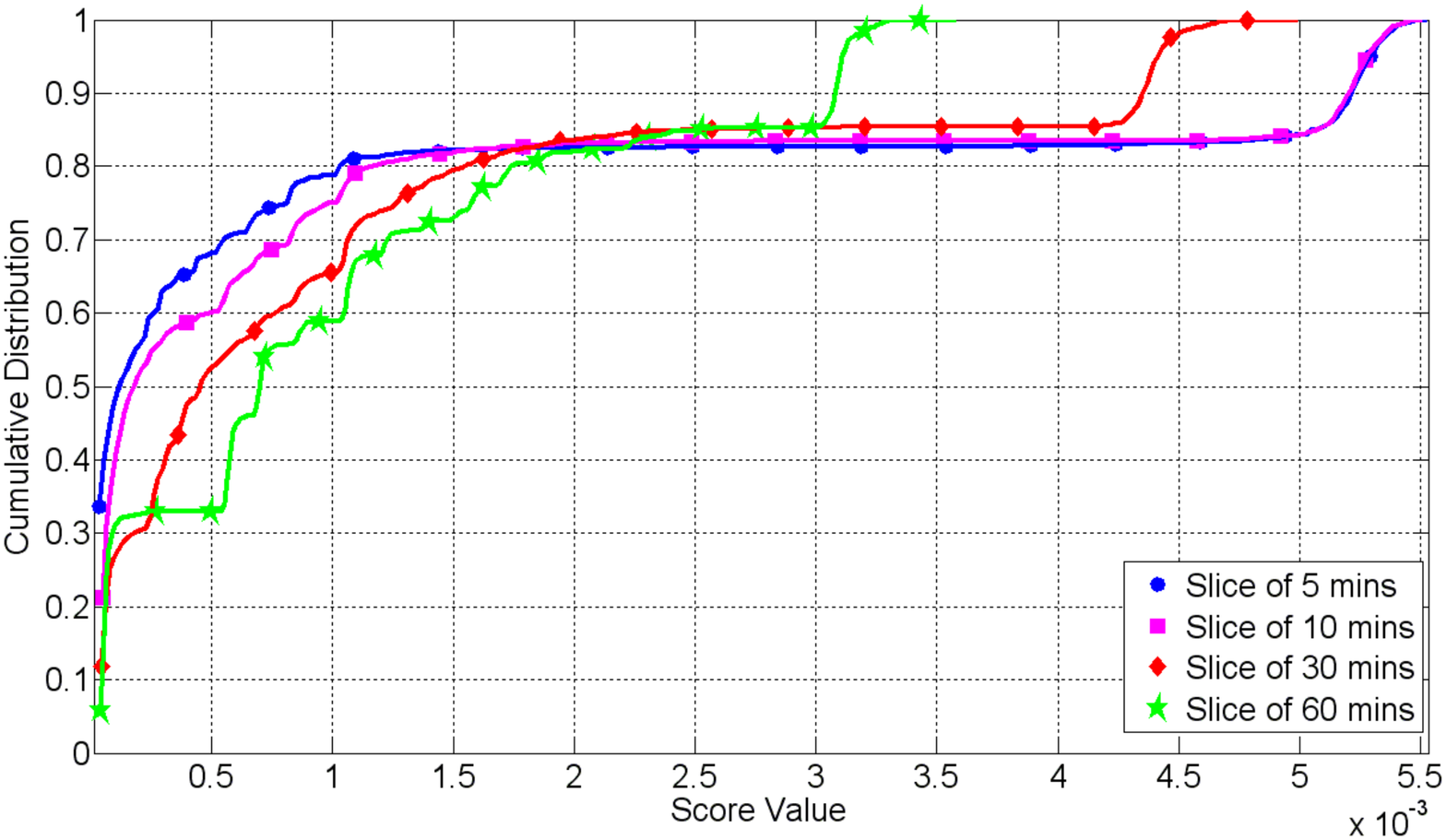}
  }
  \caption{Cumulative distribution function of Katz scores obtained from TVC model trace}
  \label{CDF_TVC}
\end{figure*}

\subsubsection{Evaluation of the link prediction technique through performance metrics}
As another evaluation step, adapted metrics are used in order to
further weigh the performance of the proposed link prediction
technique. At this step, on top of evaluating prediction of all
links, we try to focus on assessing the efficiency of our technique
in predicting new links that occurred for the first time at $T$+1
(while ignoring all previously seen links). To this end, we compute
the Area Under the ROC Curve metric (AUC metric) \cite{FAWCETT2006}
which could be considered as a good performance indicator in our
case. Thus the top scores ratio metric at $T$+1 is also considered.
To determine this metric, we compute the accurate number of links
identified through the link prediction technique. We list, for each
considered time period, the number of existing links at period
$T$+1, which we call $L$. Then, we extract the links having the $L$
highest scores and determine the number of existing links in both
sets. The evaluation metrics are computed for all traces with
different tensor slice periods in both distributed and centralized
scenarios. The results corresponding to all links prediction are
listed in Table \ref{table2} (Dartmouth Campus trace), Table
\ref{table3} (MIT Campus trace) and Table \ref{table4}. The results
corresponding to new links prediction are listed in Table
\ref{table5}, Table \ref{table6} and Table \ref{table7}(respectively
for Dartmouth Campus, MIT Campus and TVC model traces).

\begin{table}[!t]
\renewcommand{\arraystretch}{1.3}
\caption{Evaluation metrics for the prediction of all links applied
on Dartmouth Campus trace} \label{table2} \centering \scalebox{0.7}{
\begin{tabular}{|l|c|c|}
\hline
\backslashbox{\bfseries Prediction Cases}{\bfseries Metrics} & \bfseries AUC & \bfseries Top Scores Ratio at $T$+1\\
\hline\hline
Distributed Case and $t$=5 mins & 0.9850 & 2944/3267 (90.11\%)\\
Centralized Case and $t$=5 mins & 0.9844 & 2945/3267 (90.14\%)\\
\hline
Distributed Case and $t$=10 mins & 0.9817 & 2866/3340 (85.80\%) \\
Centralized Case and $t$=10 mins & 0.9826 & 2866/3340 (85.80\%) \\
\hline
Distributed Case and $t$=30 mins & 0.9360 & 2758/3832 (71.97\%) \\
Centralized Case and $t$=30 mins & 0.9324 & 2758/3832 (71.97\%) \\
\hline
Distributed Case and $t$=60 mins & 0.9153 & 2928/4270 (68.57\%) \\
Centralized Case and $t$=60 mins & 0.9069 & 2926/4270 (68.52\%) \\
\hline
\end{tabular}}

\end{table}

\begin{table}[!t]
\renewcommand{\arraystretch}{1.3}
\caption{Evaluation metrics for the prediction of all links applied
on MIT Campus trace} \label{table3} \centering \scalebox{0.7}{
\begin{tabular}{|l|c|c|}
\hline
\backslashbox{\bfseries Prediction Cases}{\bfseries Metrics} & \bfseries AUC & \bfseries Top Scores Ratio at $T$+1\\
\hline\hline
Distributed Case and $t$=5 mins & 0.9838 & 1922/2147 (89.52\%)\\
Centralized Case and $t$=5 mins & 0.9842 & 1925/2147 (89.66\%)\\
\hline
Distributed Case and $t$=10 mins & 0.9813 & 1867/2187 (85.36\%) \\
Centralized Case and $t$=10 mins & 0.9807 & 1866/2187 (85.32\%) \\
\hline
Distributed Case and $t$=30 mins & 0.9631 & 1757/2311 (76.02\%) \\
Centralized Case and $t$=30 mins & 0.9618 & 1757/2311 (76.02\%) \\
\hline
Distributed Case and $t$=60 mins & 0.9256 & 1803/2657 (67.85\%) \\
Centralized Case and $t$=60 mins & 0.9361 & 1817/2657 (68.38\%) \\
\hline
\end{tabular}}
\end{table}

\begin{table}[!t]
\renewcommand{\arraystretch}{1.3}
\caption{Evaluation metrics for the prediction of all links applied
on TVC model trace} \label{table4} \centering \scalebox{0.7}{
\begin{tabular}{|l|c|c|}
\hline
\backslashbox{\bfseries Prediction Cases}{\bfseries Metrics} & \bfseries AUC & \bfseries Top Scores Ratio at $T$+1\\
\hline\hline
Distributed Case and $t$=5 mins & 0.8851 & 717/931 (77.01\%)\\
Centralized Case and $t$=5 mins & 0.8860 & 717/931 (77.01\%)\\
\hline
Distributed Case and $t$=10 mins & 0.8409 & 750/1080 (69.44\%) \\
Centralized Case and $t$=10 mins & 0.8401 & 749/1080 (69.35\%) \\
\hline
Distributed Case and $t$=30 mins & 0.8412 & 757/1080 (70.09\%) \\
Centralized Case and $t$=30 mins & 0.8424 & 757/1080 (70.09\%) \\
\hline
Distributed Case and $t$=60 mins & 0.8388 & 755/1080 (69.90\%) \\
Centralized Case and $t$=60 mins & 0.8399 & 755/1080 (69.90\%) \\
\hline
\end{tabular}}
\end{table}

\begin{table}[!t]
\renewcommand{\arraystretch}{1.3}
\caption{Evaluation metrics for the prediction of new links applied
on Dartmouth Campus trace} \label{table5} \centering \scalebox{0.7}{
\begin{tabular}{|l|c|c|}
\hline
\backslashbox{\bfseries Prediction Cases}{\bfseries Metrics} & \bfseries AUC & \bfseries Top Scores Ratio at $T$+1\\
\hline\hline
Distributed Case and $t$=5 mins & 0.6671 & 1/144 (0.69\%)\\
Centralized Case and $t$=5 mins & 0.6518 & 1/144 (0.69\%)\\
\hline
Distributed Case and $t$=10 mins & 0.6759 & 1/184 (0.54\%) \\
Centralized Case and $t$=10 mins & 0.6913 & 1/184 (0.54\%) \\
\hline
Distributed Case and $t$=30 mins & 0.6469 & 20/684 (2.89\%) \\
Centralized Case and $t$=30 mins & 0.6269 & 24/684 (3.50\%) \\
\hline
Distributed Case and $t$=60 mins & 0.6472 & 51/1008 (5.05\%) \\
Centralized Case and $t$=60 mins & 0.6115 & 58/1008 (5.75\%) \\
\hline
\end{tabular}}
\end{table}

\begin{table}[!t]
\renewcommand{\arraystretch}{1.3}
\caption{Evaluation metrics for the prediction of new links applied
on MIT Campus trace} \label{table6} \centering \scalebox{0.7}{
\begin{tabular}{|l|c|c|}
\hline
\backslashbox{\bfseries Prediction Cases}{\bfseries Metrics} & \bfseries AUC & \bfseries Top Scores Ratio at $T$+1\\
\hline\hline
Distributed Case and $t$=5 mins & 0.6823 & 8/107 (7.47\%)\\
Centralized Case and $t$=5 mins & 0.6921 & 8/107 (7.47\%)\\
\hline
Distributed Case and $t$=10 mins & 0.7221 & 0/141 (0.00\%) \\
Centralized Case and $t$=10 mins & 0.7121 & 4/141 (2.83\%) \\
\hline
Distributed Case and $t$=30 mins & 0.6955 & 0/267 (0.00\%) \\
Centralized Case and $t$=30 mins & 0.6843 & 0/267 (0.00\%) \\
\hline
Distributed Case and $t$=60 mins & 0.6929 & 23/620 (3.70\%) \\
Centralized Case and $t$=60 mins & 0.7383 & 25/620 (4.03\%) \\
\hline
\end{tabular}}
\end{table}

\begin{table}[!t]
\renewcommand{\arraystretch}{1.3}
\caption{Evaluation metrics for the prediction of new links applied
on TVC model trace} \label{table7} \centering \scalebox{0.7}{
\begin{tabular}{|l|c|c|}
\hline
\backslashbox{\bfseries Prediction Cases}{\bfseries Metrics} & \bfseries AUC & \bfseries Top Scores Ratio at $T$+1\\
\hline\hline
Distributed Case and $t$=5 mins & 0.4954 & 0/76 (0.00\%)\\
Centralized Case and $t$=5 mins & 0.4920 & 0/76 (0.00\%)\\
\hline
Distributed Case and $t$=10 mins & 0.4758 & 2/131 (1.52\%) \\
Centralized Case and $t$=10 mins & 0.4664 & 2/131 (1.52\%) \\
\hline
Distributed Case and $t$=30 mins & 0.4730 & 2/131 (1.52\%) \\
Centralized Case and $t$=30 mins & 0.4816 & 2/131 (1.52\%) \\
\hline
Distributed Case and $t$=60 mins & 0.4583 & 2/131 (1.52\%) \\
Centralized Case and $t$=60 mins & 0.4769 & 4/131 (3.05\%) \\
\hline
\end{tabular}}
\end{table}
Regarding all links prediction results, we note, based on the high
values of AUC metric (over than 0.9 at real traces) and top scores
ratio obtained at $T$+1, that the prediction method is efficient in
predicting future links. Moreover, we note that prediction is better
when the tensor slice periods are shorter. We also observe that the
centralized and distributed matrix of scores computation achieve
similar performances. In addition, the results related to the top
scores metric attests to the fact that the prediction of all links
is efficient (at least 68\% of links are identified and this
percentage can exceed 90\% in some cases) at both centralized and
distributed scenarios. We also note that the previous observation
regarding the redundancy of the results as the tensor slice period
varies with the synthetic trace is confirmed. Indeed, The number of
existing links at $T+1$ is the same when the period $t$ is over 5
minutes. Moreover, AUC metric and top scores ratio has almost always
the same value. Nonetheless, when the prediction only concerns new
links, AUC metric values considerably decrease. This observation
presumes that the prediction is not that accurate when only new
links are considered. Given that new links are not tracked by the
tensor, their scores are low (and even null). This interpretation is
supported by the top scores ratio at $T+1$. In fact, the percentages
of identified new links are very low (no more than 8\% in the best
cases). Hence, the tensor-based link prediction technique is not
efficient when the prediction targets the occurrence of new links.
This result is also highlighted in \cite{Acar2009} and
\cite{Dunlavy2011}.

\begin{figure*}[!tb]
  \centering
  \subfigure[Area Under the ROC Curve]{
    \label{fig23}
    \includegraphics[width=0.33\textwidth,angle=270]{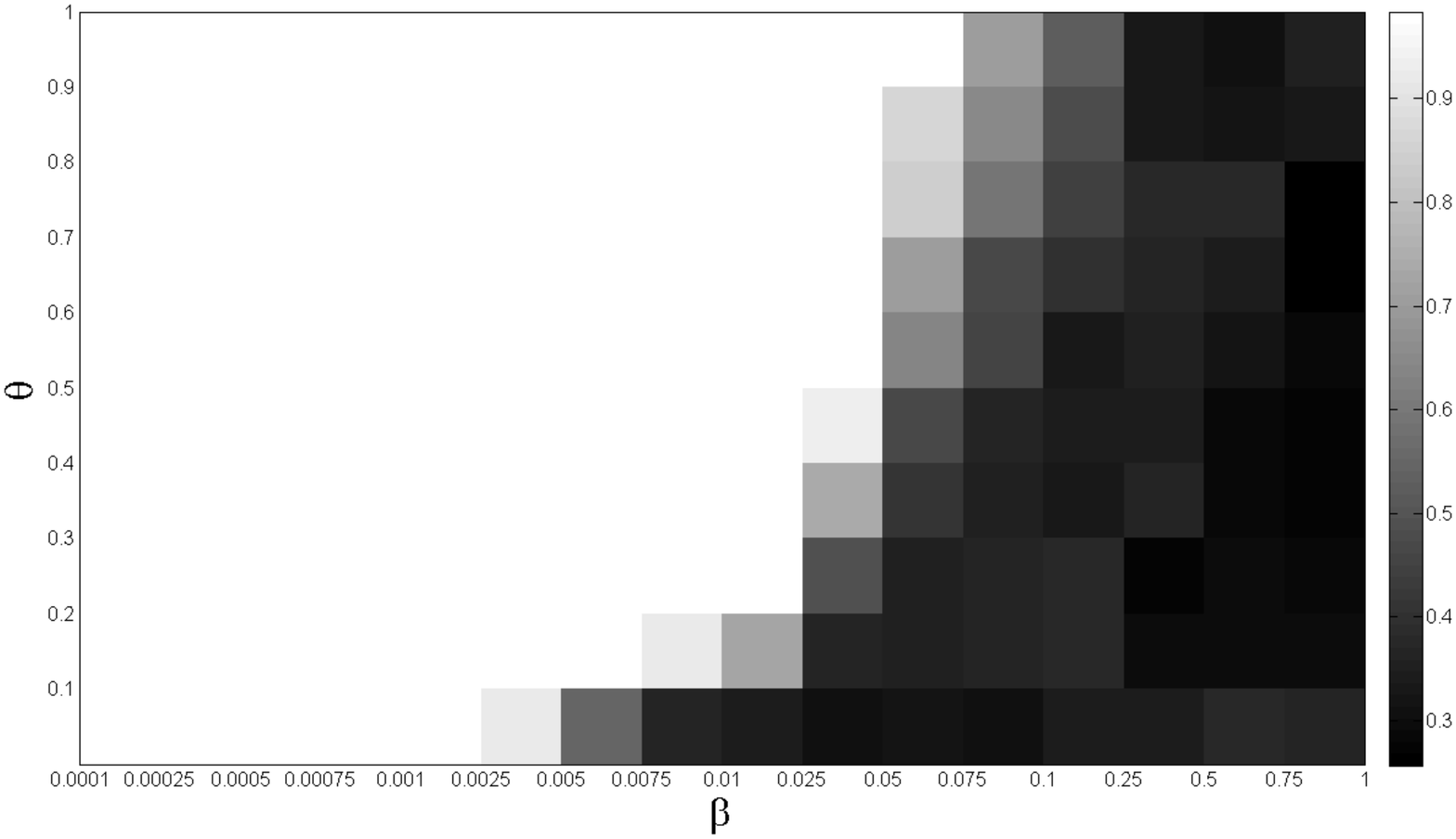}

  }
 \subfigure[Top Scores Ratio at $T$+1]{
    \label{fig24}
    \includegraphics[width=0.33\textwidth,angle=270]{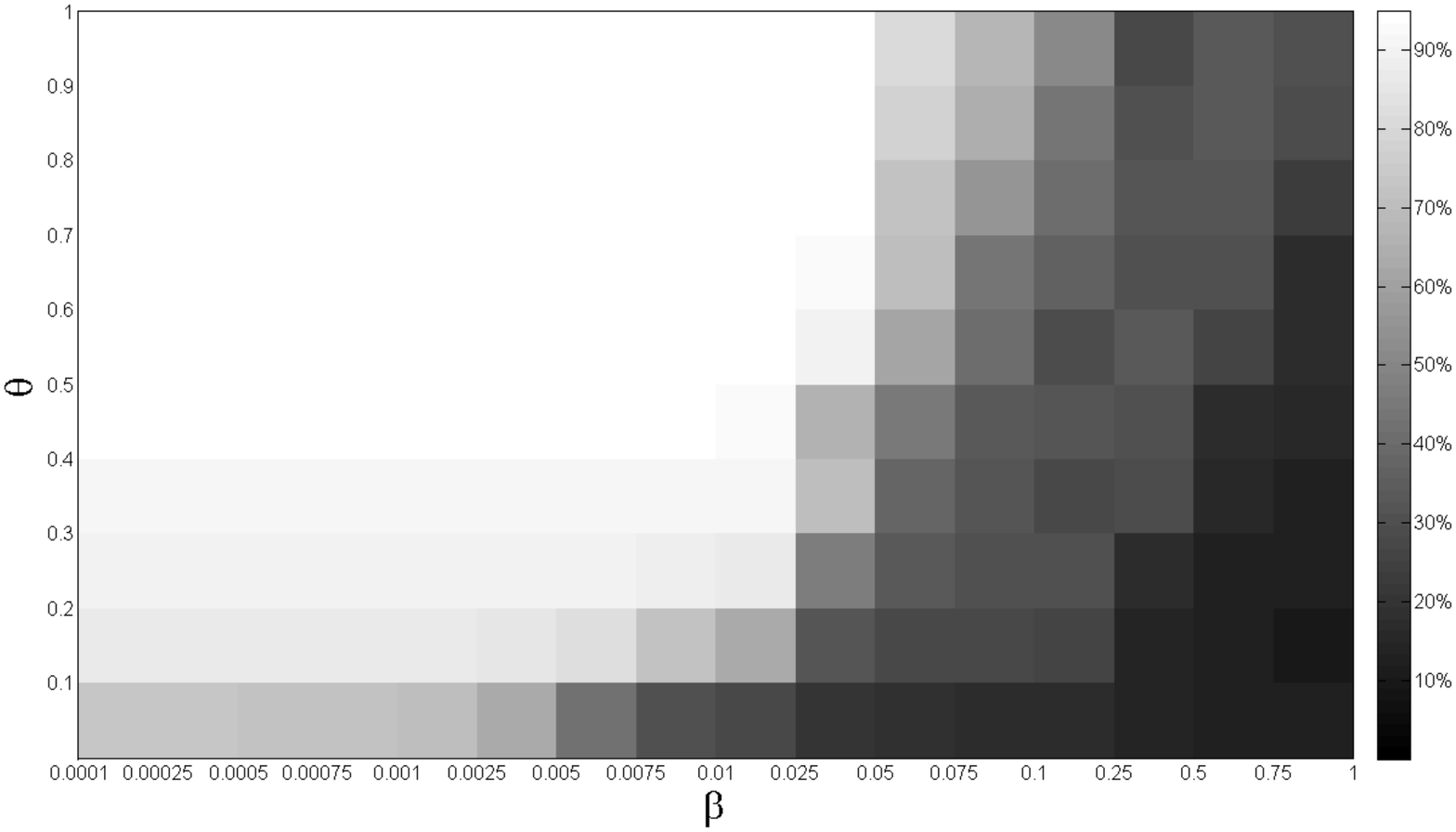}

  }
  \caption{Prediction performance of the tensor-based technique distributed version for different values of $\beta$ and $\theta$}
  \label{beta_theta}
\end{figure*}

It is also important underline the fact that our mechanism
efficiency is dependent of chosen values of $\theta$ and $\beta$. We
depict in Fig. \ref{fig23} and Fig. \ref{fig24} the top scores ratio
at $T+1$ and the AUC, respectively, obtained for different values of
$\theta$ and $\beta$. We can note that the values set to $\theta$
(i.e. 0.2) and to $\beta$ (i.e. 0.001) enables us to reach a quite
efficient level of prediction. This results are relative to a
prediction set performed on the MIT Campus trace with the
distributed version of our method (as described in the Section 4.1).

\subsubsection{Prediction Performance Comparison between the
Tensor-Based Technique and the approach of Thakur et al.}

We aim through this subsection to compare our proposal to another
similar approach (we use the distributed approach to compute the
scores). As we are designing a metric that expresses the degree of
similarity of two nodes, we choose to compare the tensor-based
technique performance to the one of the similarity metric suggested
by Thakur et al. \cite{Thakur2010}. The latter metric measures the
degree of similarity of the behaviors of two mobile nodes and the
behavior of each node is expressed by an $association$ $matrix$. The
columns of the matrix represent the possible locations that a node
can visit and the rows express time granularity (hours, days, weeks,
etc.). The dominant behavioral patterns are tracked using the
Singular Value Decomposition (SVD) \cite{Horn1990}. For more details
about the similarity metric computation, we refer the reader to
\cite{Thakur2010}.

%\begin{figure*}[!tb]
%    \centering
%    \includegraphics[width=0.6\textwidth,angle=270]{Zayani19}
%    \caption{Prediction performance comparison between the tensor-based and Thakur et al. approaches: (a) Comparison according to the Top Scores Ratio at $T$+1 metric. (b) Comparison according to the Area Under the ROC Curve metric.}
%    \label{Comparison}
%\end{figure*}

\begin{figure*}[!tb]
  \centering
  \subfigure[Comparison according to the Top Scores Ratio at $T$+1 metric]{
    \label{fig19}
    \includegraphics[width=0.25\textwidth,angle=270]{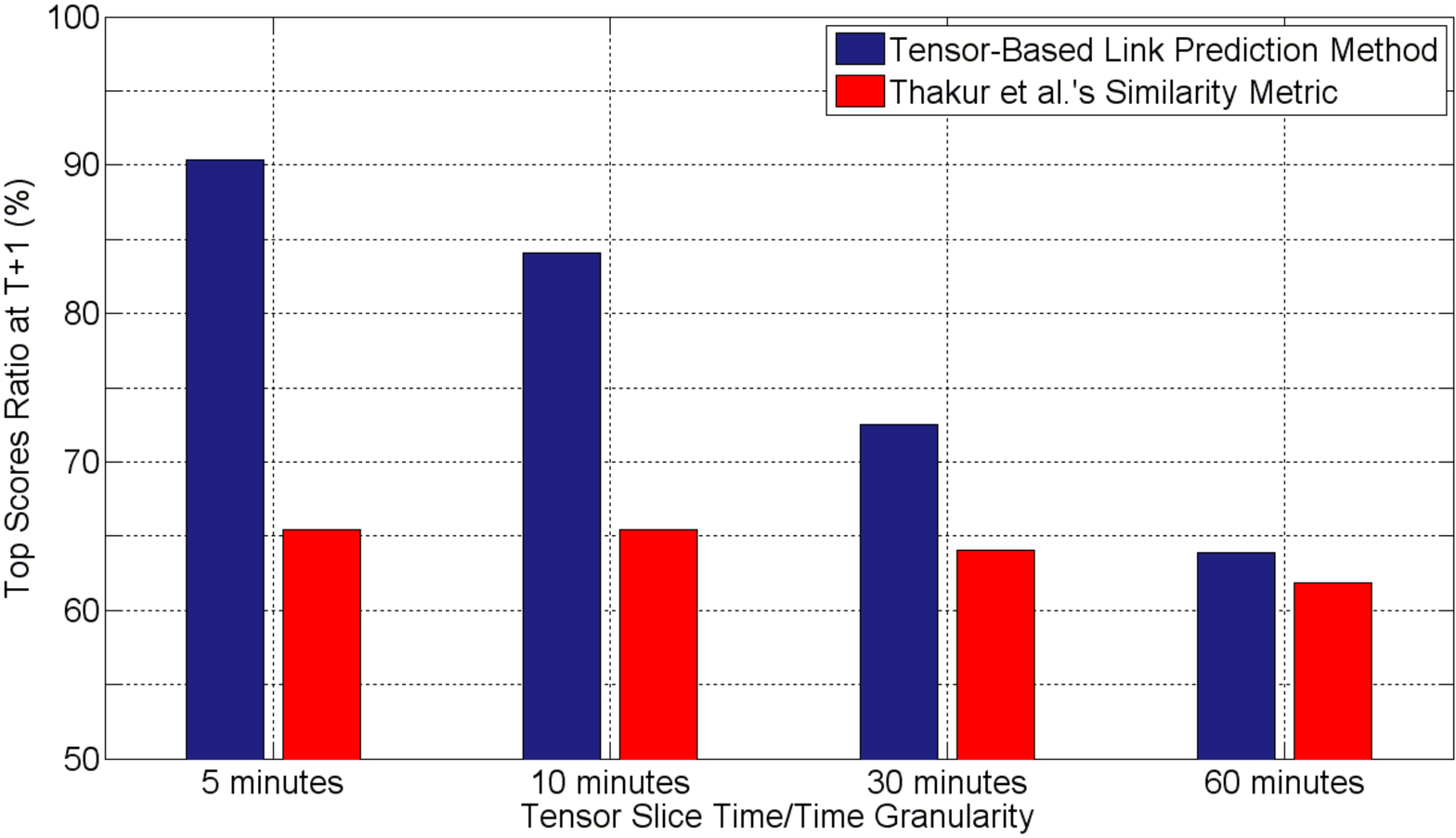}
  }\hspace{1cm}
  \subfigure[Comparison according to the Area Under the ROC Curve metric]{
    \label{fig20}
    \includegraphics[width=0.25\textwidth,angle=270]{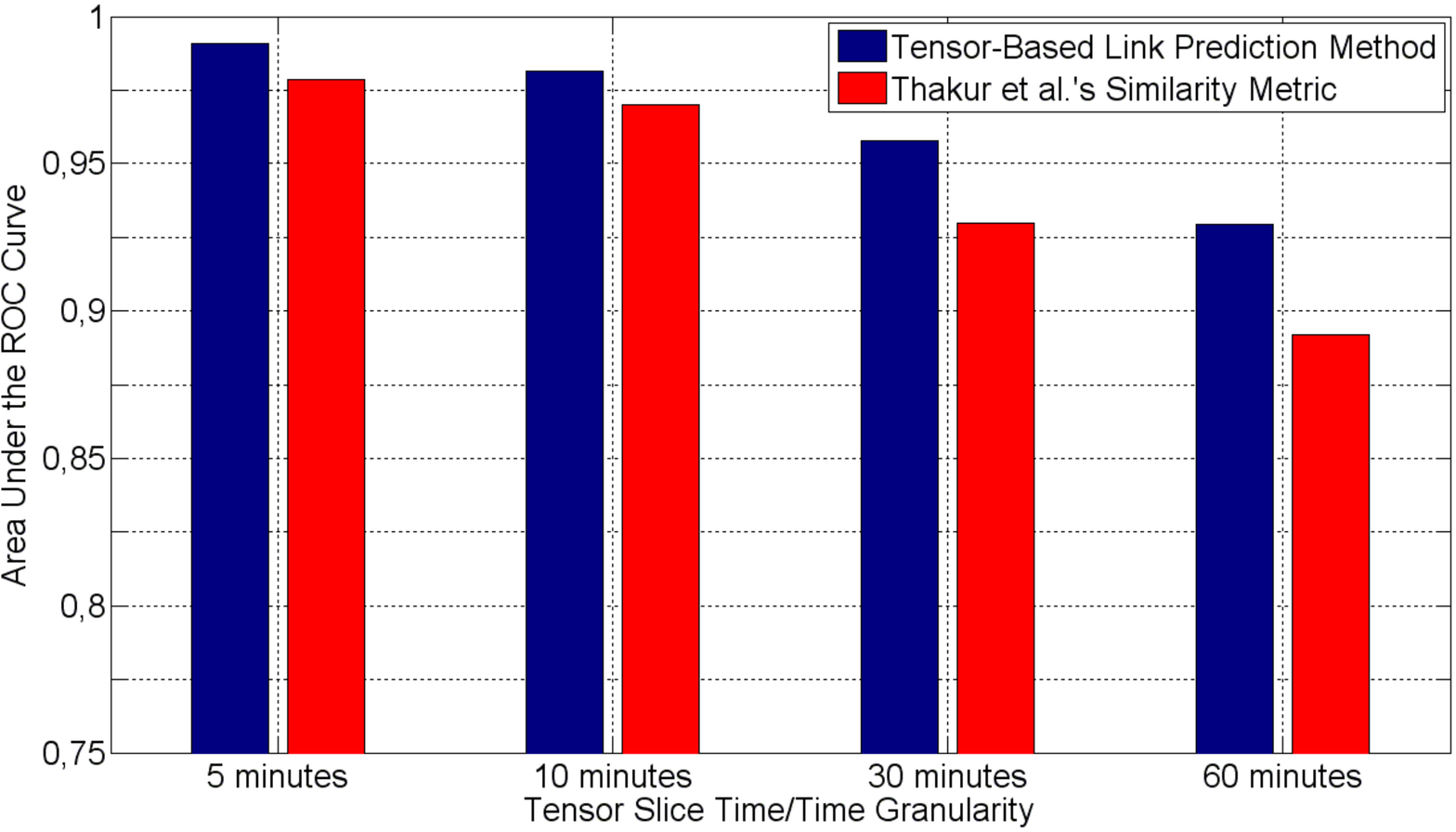}
  }\\
 \subfigure[The Top Scores Ratio at $T$+1 metric gap between the two approaches]{
    \label{fig21}
    \includegraphics[width=0.25\textwidth,angle=270]{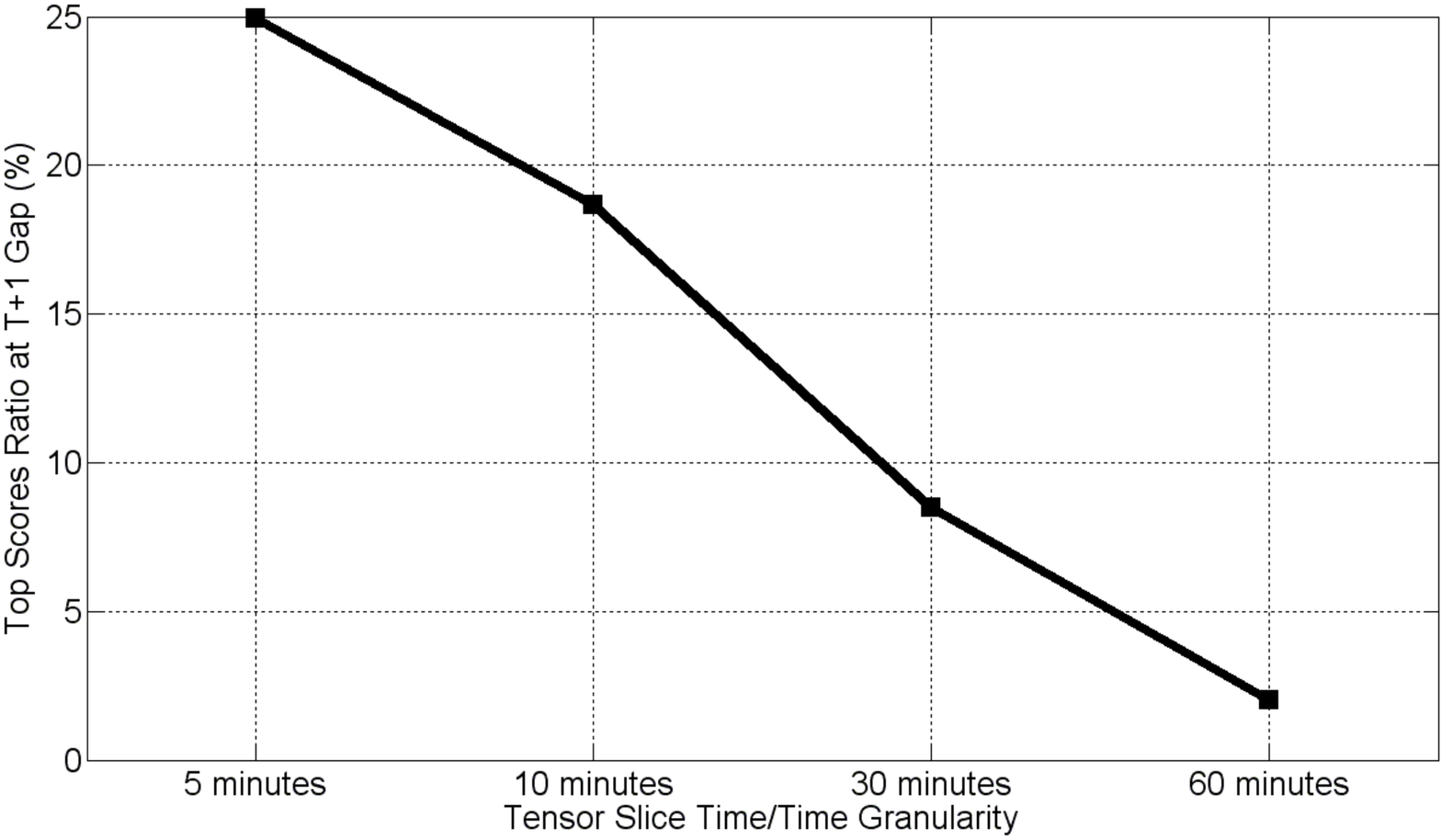}
  }\hspace{1cm}
  \subfigure[The Area Under the ROC Curve metric gap between the two approaches]{
     \label{fig22}
     \includegraphics[width=0.25\textwidth,angle=270]{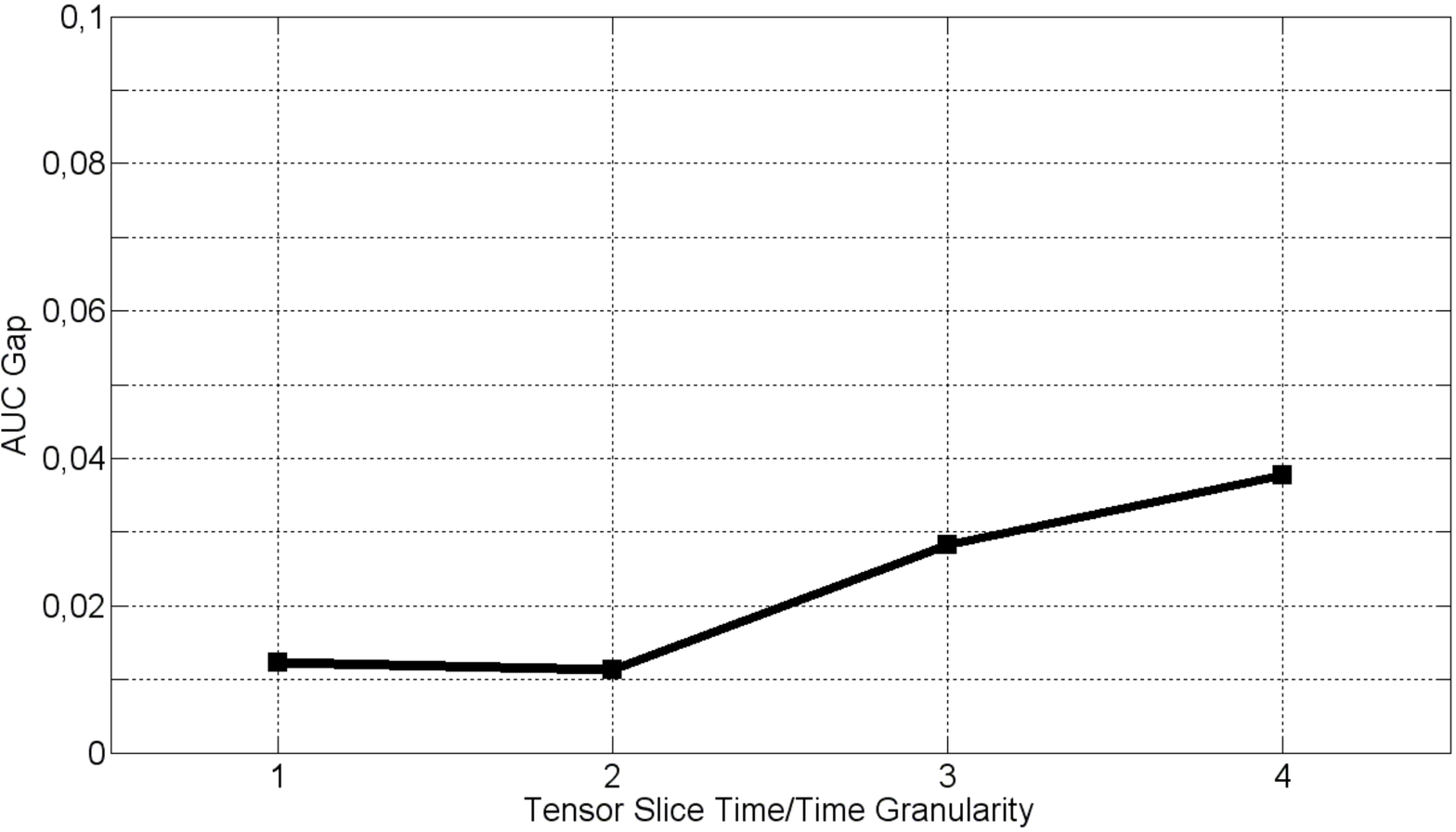}
  }
  \caption{Prediction performance comparison between the tensor-based technique and the approach of Thakur et al.}
  \label{Comparison}
\end{figure*}
%\begin{table}[!t]
%\renewcommand{\arraystretch}{1.3}
%\caption{Some tracked properties for the most accurate threshold
%used to plot the ROC curve} \label{table_comp} \centering
%\scalebox{0.7}{
%\begin{tabular}{|l|c|c|c|c|c|c|}
%\hline
%\backslashbox{\bfseries Approach}{\bfseries Metrics} & \bfseries Score & \bfseries Interval & \bfseries TP & \bfseries FP & \bfseries TN & \bfseries FN\\
%\hline\hline
%
%Tensor-Based method & 0.0029 & [0;0.005] & 1893 & 132 & 206702 & 254\\
%Similarity of Thakur et al. & 0.45 & [0;1] & 1215 & 550 & 205648 & 922\\
%\hline
%
%\end{tabular}}
%
%\end{table}
We compare the top scores ratio at $T$+1 and the area under the ROC
curve metrics and we measure them for different tensor slice times
or time granularities (5, 10, 30 and 60 minutes). For this
comparison, we use the MIT trace of 07/23/02 and track adjacency
matrices and/or association matrices from 8 a.m. to 3 p.m.. The
associated results for the top scores ratio at $T$+1 and for the
area under the ROC curve are respectively depicted in Fig.
\ref{fig19} and Fig. \ref{fig20} and the performance gap are
respectively displayed in Fig. \ref{fig21} and Fig. \ref{fig22}.

We firstly focus on the comparison according to the top scores ratio
at $T$+1. We underline that our proposal shows more efficient
prediction ability compared to Thakur et al. framework especially
when the tensor slice time/time granularity tends to be short. The
different "natures" of the metrics used in each approach explain the
results obtained for the two sets of comparison. Indeed, the measure
quantifies the similarity of nodes based on their encounters and
geographical closeness. In other words, the prediction measure cares
about contacts (or closenesses) at (around) the same location and at
the same time. Meanwhile, the similarity metric proposed by Thakur
et al. is defined as an association metric. Hence, it measures the
degree of similarity of behaviors of two mobile nodes without
necessarily seeking if they are in the same location at the same
time. As we have previously stated, the prediction performance of
the tensor-based link prediction method is better with shorter
tensor slice times. Then, with a longer tensor slice time, the
interpretation of network statistics becomes less precise. This
observation accounts for the prediction performance more comparable
for both approaches with a larger tensor slice time/time
granularity.

On the other hand the comparison based on the AUC metric, we remark
that the two approaches show quasi-similar prediction efficiency
with a slightly better performance for our proposal, mostly because
the overwhelming number of noneffective links introduce a bias in
the calculations of the AUC metric. The reduced number of occurring
links and the findings obtained for the first set of comparison
explain the little AUC gap in favor of the tensor-based link
prediction approach.
\subsection{Discussion}
In wireless networks and specifically in intermittently connected
ones, it is important to exploit social relationships that influence
nodes mobility. Taking advantage of the social aspect within these
networks could ensure a better routing strategy and therefore
improve the packets delivery rate and reduce latency. Through our
proposal, we aim to track eventual similarities between mobility
patterns of nodes and wisely exploit them for a better link
prediction.

As seen earlier, link occurrence between two nodes is more likely
when they have similar social behaviors. Then, identifying nodes
that have similar mobility pattern could help to predict effective
links between nodes in the future. The more accurate the link
prediction is the more optimized the routing scheme could get. In
fact, an efficient link prediction would help to make better
decisions in the forwarding process. For example, a node would
rather decide to postpone sending a packet to an eventual current
next hop because the link prediction scheme estimates that a better
forwarder (closer to the destination for example) is going to appear
in the immediate future. Also, link prediction could prevent buffer
overloading. Indeed, an overloaded node would rather drop a packet
if the link prediction scheme indicates that there will not be any
possible route toward the destination in the future and before the
packet's TTL expires. Through this approach, we can get quite
efficient prediction results.

As mentioned previously, the efficiency of the technique used can
exceed 90\% of identified links (with slice period equal to 5
minutes). The link prediction relies on measuring the similarity of
the mobility of nodes. Song et al. \cite{Song2010} have investigated
the limits of predictability in human mobility. Relying on data
collected from mobile phone carriers, they have found that 93\% of
user mobility is potentially predictable. The best predictability
percentage reached by our approach agrees with the conclusion of
Song et al..

We have also shown through the simulations that prediction
efficiency is similar, for a specific scenario (type of trace and
slice time period), in the case of both centralized and distributed
computation of Katz scores. As we have explained, the distributed
scheme is only able to maintain high scores (link occurrence is
likely) as nodes record neighbors at one and two hops. The seeming
lack of information does not infer on predicting effectiveness. This
observation also tallies with Acar and al. conclusion. Indeed, in
the data mining context, they have tried to make the method scalable
and proposed the Truncated Katz technique (expressed by eq. (11) in
\cite{Acar2009}). It consists in determining Katz scores replacing
the collapsed weighted tensor by a low-rank approximation one. The
results show that this latter technique retains high prediction
efficiency. Hence, restricting the scores computation on most
weighted links (in terms of recentness and duration) does not incur
dramatic consequences on prediction efficiency.

\begin{table}[!t]
\renewcommand{\arraystretch}{1.3}
\caption{Evaluation metrics for distributed prediction scenarios of
applied on MIT Campus trace} \label{table_hops} \centering
\scalebox{0.7}{
\begin{tabular}{|l|c|c|}
\hline
\backslashbox{\bfseries Prediction Cases}{\bfseries Metrics} & \bfseries AUC & \bfseries Top Scores Ratio at $T$+1\\
\hline\hline
One-hop knowledge and $t$=5 mins & 0.9747 & 1921/2147 (89.47\%)\\
Two-hops knowledge and $t$=5 mins & 0.9838 & 1922/2147 (89.52\%)\\
\hline
One-hop knowledge and $t$=10 mins & 0.9671 & 1865/2187 (85.27\%) \\
Two-hops knowledge and $t$=10 mins & 0.9813 & 1867/2187 (85.36\%) \\
\hline
One-hop knowledge and $t$=30 mins & 0.9406 & 1756/2311 (75.98\%) \\
Two-hops knowledge and $t$=30 mins & 0.9631 & 1757/2311 (76.02\%) \\
\hline
One-hop knowledge and $t$=60 mins & 0.8810 & 1789/2657 (67.33\%) \\
Two-hops knowledge and $t$=60 mins & 0.9256 & 1803/2657 (67.85\%) \\
\hline
\end{tabular}}
\end{table}

We have assumed for the computation of similarity scores using the
distributed way that nodes know their two-hop neighbors. It is
obvious that exchanging information between nodes about neighbors
causes additional overhead and consequently more solicited
resources. From this perspective, a question can be highlighted:
would the tensor-based link prediction method remain effective if
the knowledge of nodes is limited to the direct neighbors? To answer
to this question, we take into consideration the scenario where the
distributed computation of scores is based on one-hop neighboring
knowledge and we compare it to the scenario which uses the two-hops
knowledge. We use the MIT Campus trace and track the network
topology during 4 hours (i.e. the trace of 07/23/02 from 8 a.m. to
midday) and we consider different tensor slice times. The comparison
is made with the top scores ratio at $T$+1 and the AUC metrics. The
results are reported in Table \ref{table_hops}.

When the knowledge is limited to the neighbors at one hop, the
closeness only means that it exists a direct link between two nodes.
This scenario does not consider the relationships between nodes when
they are separated by multi-hops paths. The results confirm that the
prediction effectiveness is lesser. Even if the top scores ratios at
$T$+1 are close, the performance of the one-hop knowledge scenarios
are slightly worse. The AUC metric attests also that the prediction
is less efficient in such cases. In fact, considering the two-hops
knowledge generates more significative true positive rate for the
ROC curve (expressed by better top scores ratio at $T$+1) while the
false positive rate remains practically the same (due to the
overwhelming number of noneffective links). Nevertheless, the best
scenario to retain is not obvious to identify if we compare the cost
of exchanging local information between nodes to the cost of less
efficient link prediction. Future simulations and real deployments
will enable us to determine which setting is preferable to consider.

%The highest efficiency that our approach can reach matches with the
%conclusion of Song et al. \cite{Song2010} which confirms that human
%mobility is potentially predictable at 93\%.
%Regarding the satisfying results of
%our scheme, we are planning, as a future work , on designing a new
%distributed routing protocol for intermittently connected networks
%where the forwarding decision would be mainly based on the link
%prediction scheme investigated in this paper.

%\IEEEpubidadjcol

\section{Conclusion}
Human mobility patterns are mostly driven by social intentions and
correlations in the behaviors of people forming the network appear.
These similarities quantifies the correlation between the spatial
level in terms of visited locations and the temporal level regarding
mobility correlation over period of time. The knowledge about the
behavior of nodes greatly helps in improving the design of
communication protocols. Intuitively, two nodes that follow the same
social intentions over time promote the occurrence of link in the
immediate future.

In this paper, we presented a link prediction technique inspired
from data-mining and exploit it in the context of wireless networks.
Our contribution in this paper, as a new link prediction technique
for the intermittently connected wireless networks, is designed
through two major steps. First, the network topology is tracked over
several time periods in a tensor. Secondly, after collapsing the
structural information, Katz measure is computed for each pair of
nodes as a score. A high score means similar moving patterns
inferring the closeness of the nodes and indicates that a link
occurrence is likely in the future.

Through the link prediction evaluation, we have obtained relevant
results that attest the efficiency of our contribution and agree
with some findings referred in the literature. We summarize them in
the following points:
\begin{itemize}

\item The tensor-based link prediction technique is quite efficient especially when applied on real traces (Dartmouth Campus and MIT Campus traces). The result are supported by the ROC curves and the evaluation metrics (AUC and Top Scores Ratio at
$T$+1 metrics).
\item Applied on real traces, the proposed prediction technique provides more accurate results with lower tensor slice (or tensor adjacency matrices) times.
\item The prediction results with the synthetic trace (TVC model trace) confirm the lack of social
interactions. The intentions of node are only governed by the
preferred locations and do not correlate with the intentions of the
other nodes.
\item The link prediction method guarantees good performance when
prediction is applied to all links. Nevertheless, the prediction of
new links (not occurring according to statistics and by ignoring all
links seen previously) is not accurate (very low AUC and top scores
ratio at $T$+1 metrics).
\item Applying the prediction technique in a distributed way (nodes knows only their neighbors
at most at two hops) achieves similar predicting performance
compared to the use in centralized way (an entity has full-knowledge
about network structure over time).
\item The temporal tensor-based link prediction described in this paper is based on an
encounter metric which takes into account the occurring contacts at
the same location and at the same time. We provide a performance
comparison with a similar approach built around an association
similarity metric (that quantifies similarity based on preferred
locations regardless of time correlations) and show that our
proposal achieves better prediction results.

\end{itemize}

Good link prediction offers the possibility to further improve
opportunistic packet forwarding strategies by making better
decisions in order to enhance the delivery rate or limiting latency.
Therefore, it will be relevant to supply some routing protocols with
prediction information and to assess the contribution of our
approach in enhancing the performance of the network especially as
we propose an efficient distributed version of the prediction
method. The proposed technique also motivates us to inquire into
future enhancements as a more precise tracking of the behavior of
nodes and a more efficient similarity computation.

\section*{Acknowledgements}
We want to thank wholeheartedly Evrim Acar, Dimitrios Katsaros,
Walid Benameur and Rachit Agarwal for their valuable comments and
helpful advice.

%%%
%%% Biblio
%%%
%\section*{References}
%\bibliographystyle{model1-num-names}
\bibliographystyle{elsarticle-num}
\bibliography{Biblio}

\end{document}